\documentclass[a4paper,12pt]{article}
\usepackage{epsfig}
\usepackage[dvips,usenames]{color}
\usepackage{graphicx}

\newlength{\dinwidth}
\newlength{\dinmargin}
\newcommand{\spur}[1]{\not\! #1 \,}
\begin{document}
\title{Polarizations in decays $B_{u,d}\to VV$ and possible implications for
 R-parity violating supersymmetry}

\author{Ya-Dong Yang$^{1,}$\thanks{E-mail address:
yangyd@henannu.edu.cn }, Rumin Wang$^{1,2}$, G.R.Lu$^{1,2}$
\\
{\small {$^1$ \it Department of Physics, Henan Normal University,
 XinXiang, Henan 453007, P.R. China
 }}
 \\
 {\small {$^2$ \it Institute of Particle Physics, Huazhong Normal University,
  Wuhan, Hubei 430070, P.R. China}}
  }

\maketitle

\begin{abstract}

Recently  \textit{BABAR} and Belle have measured  anomalous large
transverse polarizations in some pure penguin $B \to VV$ decays,
which might be inconsistent with the Standard Model expectations. We
try to explore its implications for R-parity violating (RPV)
supersymmetry. The QCD factorization approach is employed for the
hadronic dynamics of B decays. We find  that it is possible to have
parameter spaces solving the anomaly. Furthermore, we have derived
stringent  bounds on relevant RPV couplings from the experimental
data, which is useful for further studies of RPV phenomena.

\end{abstract}

\vspace{2.0cm} \noindent {\bf PACS Numbers:  12.60.Jv,
  12.15.Mm, 12.38.Bx, 13.25.Hw}

\newpage
\section{Introduction}
The study of exclusive nonleptonic weak decays of B mesons provides
not only  good opportunities to test the standard model (SM) but
also powerful means to probe different new physics scenarios beyond
the SM. Some important results from \textit{BABAR} and Belle
collaboration on the B decays to two light vector mesons $B\to VV$
 \cite{0307026,0306007,0311017,0404029,0307014,0408093,0408017,0503013,Belle
BJ} have caught considerable theoretical interests and  have been
extensively studied   very recently
 \cite{kagan1,grossman,Colangelo,alvarez,kagan,ladisa,cheng}.

In $B\rightarrow VV$ decay modes,  both longitudinal and transverse
polarization states are possible. It has been known to us for many
years that longitudinal polarization fraction is dominant in the SM
when both the final vector mesons are light \cite{ali}. In the heavy
quark limit, dynamics for the situation is analyzed clearly by Kagan
\cite{kagan2} and  Grossman \cite{grossman}. Recent \textit{BABAR}
and Belle measurements of decay modes $B \to \rho^+ \rho^-, \rho^0
\rho^+$ and $\rho^0 K^{*+}$ have confirmed this prediction as the
final states are dominated by the longitudinal polarization.
However, for the observed $B\rightarrow \phi K^*$ and $\rho^+
K^{*0}$ decays, the transverse polarization fractions are measured
to be anomalous as large as $50\%$, which has posed a challenge to
theoretical explanation.

Recent studies have provided some possible resolutions to the
anomaly. Kagan \cite{kagan} has shown it could be solved in the SM
by increasing the nonfactorizable contributions of annihilation
diagrams, which are formally power suppressed and depend on some
poorly known nonperturbative QCD parameters. Later on, using a
different factorization approach, Li and Mishima \cite{lihn} have
found that the annihilation contribution is not sufficient to lower
the longitudinal fraction in $B\to\phi K^*$ down to $\sim 50\%$,
although it could help to alleviate the anomaly. It is also shown
that the anomaly might be due to large charming penguin
contributions and final-state-interactions(FSI) by Colangelo
$et~al.$ \cite{Colangelo} and Ladisa $et~al.$ \cite{ladisa}.
However, Cheng et al. \cite{cheng} have found the FSI effects not
able to fully account for the anomaly.   Hou and Nagashima
\cite{hou} have given a model where the transverse $\phi$ descends
from the transverse gluon from $b\to s $g$^*$. However, it should be
noted that $\phi$ must couple to at least three gluons to neutralize
color and conserve relevant quantum number. In this paper, we will
explore the opportunity whether the RPV supersymmetry (SUSY) could
provide a solution to the polarization anomaly and what kinds of
constraints could be derived.

The possible appearance of RPV couplings \cite{SUSY}, which will
result in lepton and baryon number conservation, has gained full
attention in searching for SUSY \cite{report,allanach}. The rich
phenomenology implied by RPV couplings in B decays have been
extensively discussed in \cite{RPVstudy, cskim, desh, hexg, yd}. In
this paper, we extend these studies to decays $B\to VV$. We use the
QCD factorization(QCDF) framework \cite{BBNS} for hadronic dynamics.
We show that the  polarization anomaly  could be solved in the
presence of RPV couplings, due to the appearance of $\bar{q}(V\pm
A)q \bar{b}(V+A)s$ interactions. Compared with the matrix element of
$\langle K^{*}|\bar{b}(V-A)s|B\rangle$, the axial parts in $\langle
K^{*}|\bar{b}(V+A)s|B\rangle$ have opposite sign, hence induce
polarization phenomena different from the SM. Moreover, using the
recent experimental data for branching ratios and
longitudinal-polarization fractions, we obtain limits on the
relevant RPV couplings.

This paper is organized as follows: In Sec. II, we calculate the
$B\to VV$ decay amplitudes which contain the SM part and RPV effects
using the QCDF approach. In Sec. III, we list the theoretical input
parameters used in our analysis. Section IV is devoted to the
numerical results and discussions, we also display the allowed
regions of the parameter space that satisfy all experimental data.
Finally, our summary is presented  in Sec. V.

\section{Amplitudes  for $B\to VV$ decays in QCD factorization approach }
Approaches for calculating amplitudes of B charmless nonleptonic
decays always invoke factorization frameworks. Once factorization
framework chosen, one can start from the $\Delta B=1$ effective
Hamiltonian of the underlying full electroweak (EW) theory at the
renormalization scale $\mu \sim m_b $, which can be obtained from
the full EW theory by integrating out heavy particles with mass
larger than $m_b$ using the renormalization group equation. The QCDF
\cite{BBNS} developed by Beneke, Buchalla, Neubert and Sachrajda is
a powerful approach, which will be employed in this paper. Details
of QCDF could be found in papers \cite{BBNS}.

\subsection{ The decay amplitudes  in the SM }
  In the SM, the low energy effective Hamiltonian for
  the $\Delta B=1$ transition has the form \cite{coeff}
 \begin{equation}
 {\cal
 H}^{SM}_{eff}=\frac{G_f}{\sqrt{2}}[v_u(C_1O_1^u+C_2O_2^u)+v_c(C_1O_1^c+C_2O_2^c)
 -v_t(\sum_{i=3}^{10}C_iO_i+C_{7\gamma}O_{7\gamma}+C_{8g}O_{8g})]+H.c.,
 \end{equation}
 with the effective operators given by
 \begin{equation}
\begin{array}{llllll}
O_1^u&=&(\bar{b}_\alpha u_\alpha)_{V-A}(\bar{u}_\beta
q_\beta)_{V-A},&
O_2^u&=&(\bar{b}_\alpha u_\beta)_{V-A}(\bar{u}_\beta q_\alpha)_{V-A},\\
O_1^c&=&(\bar{b}_\alpha c_\alpha)_{V-A}(\bar{c}_\beta
q_\beta)_{V-A},&
O_2^c&=&(\bar{b}_\alpha c_\beta)_{V-A}(\bar{c}_\beta q_\alpha)_{V-A},\\
O_{3(5)}&=&(\bar{b}_\alpha
q_\alpha)_{V-A}\sum\limits_{q'}(\bar{q}'_\beta
q'_\beta)_{V-A(V+A)},& O_{4(6)}&=&(\bar{b}_\alpha q_\beta)
_{V-A}\sum\limits_{q'}(\bar{q}'_\beta q'_\alpha)_{V-A(V+A)},\\
O_{7(9)}&=&\frac{3}{2}(\bar{b}_\alpha
q_\alpha)_{V-A}\sum\limits_{q'}e_{q'}(\bar{q}'_\beta q'_\beta)
_{V+A(V-A)},& O_{8(10)}&=&\frac{3}{2}(\bar{b}_\alpha q_\beta)
_{V-A}\sum\limits_{q'}e_{q'}(\bar{q}'_\beta q'_\alpha)_{V+A(V-A)},\\
O_{7\gamma}&=&\frac{e}{8\pi^2}m_b\bar{b}_\alpha\sigma^{\mu\nu}
 (1+\gamma_5)q_{\alpha}F_{\mu\nu},&
O_{8g}&=&\frac{g}{8\pi^2}m_b\bar{b}_\alpha\sigma^{\mu\nu}
 (1+\gamma_5)T_{\alpha\beta}^{a}q_{\beta}G_{\mu\nu}^{a}.
\end{array}
\end{equation}
Where $v_p=V^*_{pb}V_{pq}$ are the Cabibbo-Kobayashi-Maskawa (CKM)
\cite{CKMdef} factors, $C_i({\mu})$ are the effective Wilson
coefficients and could be found in Ref. \cite{coeff}; $\alpha$,
$\beta$ are the SU(3) color indices, $(\bar{q}_1q_2)_{V\pm
A}\equiv\bar{q}_1\gamma^\mu (1\pm \gamma_5)q_2$, $q=d,s$ and
$q'=u,d,s,c,b$.

In order to calculate the decay amplitudes and branching ratios
for $B\to VV$ decays, we need the hadronic matrix element of the
local four fermion operators
 \begin{equation}
  \langle V_1(\lambda_1)V_2(\lambda_2)|(\bar{q}_2q_3)_{V\pm A}
(\bar{b}q_1)_{V-A}|B\rangle,
 \end{equation}
where $\lambda_1$, $\lambda_2$ are the helicities of the final-state
vector mesons $V_1$ and $V_2$ with four-momentum $p_1$ and $p_2$,
respectively. In the rest frame of $B$ system, since the $B$ meson
has spin zero,  two vectors have the same helicity therefore three
polarization states are possible, one longitudinal (L) and two
transverse, corresponding to helicities $\lambda=0$ and
$\lambda=\pm$ ( here $\lambda_1=\lambda_2=\lambda$). We use
$X^{(BV_1,V_2)}$ to denote the factorizable amplitude for different
chiral currents  with the vector meson $V_2$ being factorized out.
Under the naive factorization (NF) approach, we can express it as
\begin{eqnarray}
 X^{(BV_1,V_2)}\equiv \langle V_2|(\bar{q}_2\gamma_\mu(1-a\gamma_5)q_3)
 |0\rangle
 \langle V_1|(\bar{b}\gamma^\mu(1-\gamma_5)q_1)|B \rangle,
 \end{eqnarray}
here $a=+1$ or $-1$ corresponding to $(\bar{q}_2 q_3 )_{(V-A)}$ or
$(\bar{q}_2 q_3 )_{(V+A)}$ current, respectively.

 We write the
factorized matrix elements in term of the decay constant and form
factors \cite{BallZwicky}
\begin{eqnarray}
\langle V(p,\varepsilon^{\ast})|\bar{q}\gamma_{\mu}q'|0\rangle&=&
f_{V}m_V\varepsilon_{\mu}^{\ast},\\
\langle
V(p,\varepsilon^{\ast})|\bar{b}\gamma_{\mu}q|B(p_{B})\rangle &=&
\frac{2V^{BV}(q^2)}{m_B+m_V}
\epsilon_{\mu\nu\alpha\beta}\varepsilon^{\ast\nu}p_B^{\alpha}p^{\beta}
,\\
\langle V(p,\varepsilon^{\ast})|\bar{b}\gamma_{\mu}\gamma_5q|B
(p_{B})\rangle
&=&i\left[\varepsilon_{\mu}^\ast(m_B+m_V)A_1^{BV}(q^2)
-(p_B+p)_{\mu}({\varepsilon^\ast}\cdot{p_B})\frac{A_2^{BV}(q^2)}
{m_B+m_V}\right]\nonumber \\
&&-iq_{\mu}({\varepsilon^\ast}\cdot{p_B})\frac{2m_V}{q^2}
[A_3^{BV}(q^2)-A_0^{BV}(q^2)],
\end{eqnarray}
where $p_B (m_{B})$ is the four-momentum (mass) of the $B$ meson,
$m_{V_1} (\varepsilon_1^{\ast})$ and $m_{V_2}
(\varepsilon_2^{\ast})$ are the masses (polarization vectors) of the
vector mesons $V_1$ and $V_2$, respectively,
 $q=p_B-p$ is the transferred momentum, and the form factors obey the
following exact relations \cite{PV}
\begin{eqnarray}
A_3(0)&=&A_0(0)\nonumber,\\
A_3^{BV}(q^2)&=&\frac{m_B+m_V}{2m_V}A_1^{BV}(q^2)-\frac{m_B-m_V}{2m_V}A_2^{BV}(q^2).
\end{eqnarray}
One has the factorizable  $B \to V_1 V_2$ amplitude
\begin{eqnarray}
X^{(B V_1,V_2)}&=&-i f_{V_2}m_{V_2}\biggl[
(\varepsilon_1^{\ast}\cdot\varepsilon_2^{\ast})
(m_{B}+m_{V_1})A_1^{BV_1}(m_{V_2}^2)\biggr. \nonumber\\
&&-(\varepsilon_1^{\ast}\cdot p_B)(\varepsilon_2^{\ast}\cdot
p_B)\frac{2A_2^{BV_1}(m_{V_2}^2)}{m_{B}+m_{V_1}}
\left.+i\epsilon_{\mu\nu\alpha\beta}\varepsilon_2^{\ast\mu}
\varepsilon_1^{\ast\nu}p_B^{\alpha}p_1^{\beta}
\frac{2V^{BV_1}(q^2)}{m_{B}+m_{V_1}}\right].
\end{eqnarray}
We assume the $V_1$ ($V_2$) meson flying in the minus(plus)
z-direction carrying the momentum $p_1$ ($p_2$), Using the sign
convention $\epsilon^{0123}=-1$, we get
\begin{eqnarray}
 X^{(BV_1,V_2)}&=&\left \{\begin{array}{ll}\frac{i
f_{V_2}}{2m_{V_1}}\left[(m^2_{B}-m_{V_1}^2-m_{V_2}^2)(m_B+m_{V_1})
A_1^{BV_1}(m_{V_2}^2)-\frac{4m_{B}^2p_c^2}{m_{B}+m_{V_1}}
A_2^{BV_1}(m_{V_2}^2)\right]\equiv
h_0,
\\i
 f_{V_2}m_{V_2}[(m_{B}+m_{V_1})A_1 ^{BV_1}(m_{V_2}^2)\mp
\frac{2m_{B}p_c}{m_{B}+m_{V_1}}V ^{BV_1}(m_{V_2}^2)]\equiv
h_{\pm},
 \end{array}
 \right.
 \end{eqnarray}
 where $h_0$ for $\lambda=0$ and $h_\pm$ for
$\lambda=\pm$.

The QCDF approach \cite{BBNS} allows us to compute the
nonfactorizable corrections to the hadronic matrix elements $\langle
V_1 V_2|O_i|B\rangle$ in the heavy quark limit. The nonfactorizable
corrections can be normalized to the factorizable amplitudes, so
that they enter  the effective parameters $a_i^\lambda$  as
$\alpha_s $ corrections. They are calculated from the vertex
corrections, hard spectator interactions, and QCD penguin-type
contributions.

In QCDF approach, light-cone distribution amplitudes (LCDAs) play an
essential role. The LCDAs of the light vector meson are given in
\cite{PV}. In general, the light-cone projection operator for vector
mesons in momentum space can be divide into two parts
\begin{equation}
{\cal M}^V={\cal M}_{\parallel}^V+{\cal M}_{\perp}^V.
\end{equation}
The longitudinal projector and the transverse projector are given by
\cite{PV,MT}
\begin{eqnarray}
{\cal M}^V_{\parallel}&=& \frac{f_V}{4}\frac{m_V
(\varepsilon^{\ast} \cdot n_+)}{2}\spur{n_-} \Phi^V_{\parallel}
(u)+\frac{f^\perp_Vm_V}{4}\frac{m_V
(\varepsilon^{\ast} \cdot n_+)}{2E}\left[-\frac{i}{2}
\sigma_{\mu\nu}n^\mu_-n^\nu_+h^{(t)V}_\parallel(u)\right.\nonumber\\
&&\left.-iE\int_0^u
dv(\phi^V_\perp(v)-h_\parallel^{(t)V}(v))
\sigma_{\mu\nu}n_-^\mu\frac{\partial}{\partial
k_{\perp \nu}}+\frac{h'^{(s)V}_\parallel(u)}{2}
\right]\Bigg|_{k=up'},\label{paral}
\end{eqnarray}
\begin{eqnarray*}
M^V_{\delta\alpha\perp}&=&\frac{f^\perp_V}{4}E\rlap/\varepsilon^*_\perp
\rlap/n_-\Phi^V_\perp(u)
+\frac{f_Vm_V}{4}\Biggl\{\rlap/\varepsilon^*_\perp
\mbox{g}^{(v)V}_\perp(u) -E\int_0^u
dv(\Phi^V_\parallel(v)-\mbox{g}_\perp^{(v)V}(v))\rlap/n_-
\varepsilon^*_{\perp\mu}\frac{\partial}{\partial k_{\perp \mu}}
\hspace{5.3cm} \nonumber\\
&&+i\epsilon_{\mu \nu \rho \sigma}\varepsilon^{* \nu}_\perp n_-^\rho
\gamma^\mu
\gamma_5\left[n_+^\sigma\frac{\mbox{g}_\perp^{'(a)V}(u)}{8}
-E\frac{\mbox{g}_\perp^{(a)V}(u)}{4}\frac{\partial}{\partial
k_{\perp \sigma}}\right]
 \Biggl\}\Bigg|_{k=up'}.
\end{eqnarray*}
Here we suppose the vector meson moving in the $n_-$ direction,
$n_{\pm}=(1,0,0,\pm 1)$ is the light-cone vectors, $u$ is the
light-cone momentum fraction of the quark in the vector meson, $f_V$
and $f^{\perp}_V$ are vector and tensor decay constants,
respectively, and E is the energy of the vector meson in the B rest
system. In the main body of the paper we neglect power-suppressed
higher-twist effects, i.e. we identify the meson momentum $p'\equiv
En_-$ and set $\varepsilon^*\cdot n_-=0$.

In the heavy quark limit, the light-cone projector for $B$ meson can
be expressed as \cite{bbnspk}
 \begin{equation}
 {\cal M}^B=-\frac{i f_B m_B}{4}\left[(1+\spur{v}
)\gamma_5 \left\{\Phi_1^B (\xi) +\spur{n_-} \Phi_2^B
(\xi)\right\}\right],
 \label{projector}
 \end{equation}
with $v=(1,0,0,0)$ and the normalization conditions
 \begin{equation}
 \int_{0}^{1}{\rm d}{\xi}\Phi_1^B (\xi)=1,~~~~~\int_{0}^{1}{\rm
d}{\xi}\Phi_2^B (\xi)=0,
 \end{equation}
where $\xi$ is the momentum fraction of the spectator quark in the
$B$ meson.

The coefficients of the flavor operators $a_i^\lambda$ which contain
next-to-leading order coefficient and $\mathcal{O}(\alpha_s)$ hard
scattering corrections can be written as follows:
\begin{eqnarray}
a_1^\lambda &=&C_1+\frac{C_2}{N_C}+\frac{\alpha_s}{4\pi}
\frac{C_F}{N_C}C_2\Big[f_I^\lambda(1)+f_{II}^\lambda(1)\Big],\nonumber\\
a_2^\lambda &=&C_2+\frac{C_1}{N_C}+\frac{\alpha_s}{4\pi}
\frac{C_F}{N_C}C_1\Big[f_I^\lambda(1)+f_{II}^\lambda(1)\Big],\nonumber\\
a_3^\lambda &=&C_3+\frac{C_4}{N_C}+\frac{\alpha_s}{4\pi}
\frac{C_F}{N_C}C_4\Big[f_I^\lambda(1)+f_{II}^\lambda(1)\Big],\nonumber\\
a_4^\lambda &=&C_4+\frac{C_3}{N_C}+\frac{\alpha_s}{4\pi}
\frac{C_F}{N_C}C_3\Big[f_I^\lambda(1)+f_{II}^\lambda(1)\Big]
+\frac{\alpha_s}{4\pi}\frac{C_F}{N_C} \left\{-C_1
\Big[\frac{v_u}{v_t}G^\lambda (s_u)+\frac{v_c}{v_t}
 G^\lambda (s_c)\Big]\right.\nonumber\\&&+C_3\Big[G^\lambda
(s_q)+G^\lambda(s_b)\Big]+(C_4+C_6)\sum_{q'=u}^b \Big[G^\lambda
(s_{q'})-\frac{2}{3}\Big] +\frac{3}{2}C_9\Big[e_qG^\lambda
(s_q)+e_bG^\lambda(s_b)\Big]\nonumber\\&&+\left.
\frac{3}{2}(C_8+C_{10})
\sum_{q'=u}^be_{q'}\Big[G^\lambda(s_{q'})-\frac{2}{3}\Big]+C_{8g}
G_g^\lambda \right\}, \nonumber\\
a_5^\lambda &=&C_5+\frac{C_6}{N_C}-\frac{\alpha_s}{4\pi}
\frac{C_F}{N_C}C_6\Big[f_I^\lambda(-1)+f_{II}^\lambda(-1)\Big],\nonumber\\
a_6^\lambda &=&C_6+\frac{C_5}{N_C},\nonumber\\
a_7^\lambda &=&C_7+\frac{C_8}{N_C}-\frac{\alpha_s}{4\pi}
\frac{C_F}{N_C}C_8\Big[f_I^\lambda(-1)+f_{II}^\lambda(-1)\Big]
-\frac{\alpha_e}{9\pi}N_CC^\lambda_e,\nonumber\\
a_8^\lambda &=&C_8+\frac{C_7}{N_C},\nonumber\\
a_9^\lambda &=&C_9+\frac{C_{10}}{N_C}+\frac{\alpha_s}{4\pi}
\frac{C_F}{N_C}C_{10}\Big[f_I^\lambda(1)+f_{II}^\lambda(1)\Big]
-\frac{\alpha_e}{9\pi}N_CC^\lambda_e,\nonumber\\
a_{10}^\lambda &=&C_{10}+\frac{C_9}{N_C}+\frac{\alpha_s}{4\pi}
\frac{C_F}{N_C}C_9\Big[f_I^\lambda(1)+f_{II}^\lambda(1)\Big]
-\frac{\alpha_e}{9\pi}C^\lambda_e, \label{coeff}
\end{eqnarray}
where $C_F=(N_C^2-1)/(2 N_C)$, $s_i=m_i^2/m_b^2$, and $N_C=3$ is
the number of colors. The superscript $\lambda$ denotes the
polarization of the vector meson.

In Eq.(\ref{coeff}), $f_I^{\lambda}(\pm 1)$ contain the
contributions from the vertex corrections, and given by
 \begin{eqnarray}
 &&f_I^0(a)=-12\ln \frac{\mu}{m_b}-18+6(1-a)+\int_0^1{\rm d}u
\Phi_{\parallel}^{V_2} (u)\left(3\frac{1-2u}{1-u}\ln u-3i \pi\right),\\
 &&f_I^{\pm}(a)=-12\ln \frac{\mu}{m_b}-18+6(1-a)+\int_0^1{\rm d}u
\left(\mbox{g}_{\perp}^{(v){V_2}} (u)\pm \frac{a
\mbox{g}_{\perp}^{\prime(a){V_2}}
(u)}{4}\right)\left(3\frac{1-2u}{1-u}\ln u-3i \pi\right).\nonumber
 \end{eqnarray}

For hard spectator scattering contributions, we  use the notation
that ${V_1}$ is the recoiled meson and $V_2$ is the emitted meson,
explicit calculations for $f_{II}^\lambda$ yield
\begin{eqnarray}
f^{0}_{II}(a)&=&\frac{4\pi^2}{N_C}\frac{if_{B}
f_{V_1}f_{V_2}}{h_0}\int_0^1{\rm d}\xi \frac{\Phi_1^B
(\xi)}{\xi}\int_0^1{\rm d}v \frac{\Phi_{\parallel}^{V_1}
(v)}{\bar{v}}\int_0^1{\rm d}u \frac{\Phi_{\parallel}^{V_2}
(u)}{u},\nonumber\\
f_{II}^{\pm}(a)&=&-\frac{4
\pi^2}{N_C}\frac{2if_{B}f^{\perp}_{V_1}f_{V_2}m_{V_2}}{m_{B}h_{\pm}}
(1\mp1)\int_0^1{\rm d}\xi \frac{\Phi_1^B (\xi)}{\xi}\int_0^1{\rm
d}v \frac{\Phi_{\perp}^{V_1} (v)}{\bar{v}^2}\nonumber
\\&&\times\int_0^1{\rm d}u \left(\mbox{g}_{\perp}^{(v){V_2}}
(u)-\frac{a\mbox{g}_{\perp}^{\prime(a){V_2}} (u)}{4}\right)+\frac{4
\pi^2}{N_C}\frac{2if_{B}f_{V_1}f_{V_2}m_{V_1}m_{V_2}}{m_{B}^2h_{\pm}}
\int_0^1{\rm d}\xi \frac{\Phi_1^B (\xi)}{\xi}\nonumber
\\&&\times\int_0^1{\rm d}v {\rm d}u\left(\mbox{g}_{\perp}^{(v){V_1}} (v)
\pm\frac{\mbox{g}_{\perp}^{\prime(a){V_1}}
(v)}{4}\right)\left(\mbox{g}_{\perp}^{(v){V_2}} (u)\pm \frac{a
\mbox{g}_{\perp}^{\prime(a){V_2}} (u)}{4}\right)\frac{u+\bar v}
{u{\bar v}^2}, \label{hard}
\end{eqnarray}
with $\bar{v}=1-v$. In Eq.(\ref{hard}), when the asymptotical form
for the vector meson LCDAs adopted, there will be infrared
divergences  in $f_{II}^{\pm}$. As in \cite{bbnspk,ykc}, we
introduce a cutoff of order $\Lambda_{QCD}/m_b$ and take
$\Lambda_{QCD}=0.5$ GeV as our default value.

The contributions of the QCD penguin-type diagrams can be
described by the functions
\begin{eqnarray}
G^0 (s)&=&\frac{2}{3}-\frac{4}{3}\ln \frac{\mu}{m_b}+4\int_0^1{\rm
d}u ~\Phi_{\parallel}^{V_2}
(u) \mbox{g}(u,s),\\
G^{\pm}(s)&=&\frac{2}{3}-\frac{2}{3}\ln
\frac{\mu}{m_b}+2\int_0^1{\rm d}u
~(\mbox{g}_{\perp}^{(v){V_2}}(u)\pm
\frac{\mbox{g}_{\perp}^{\prime(a){V_2}}(u)}{4}) \mbox{g}(u,s),
\end{eqnarray}
with the function g$(u,s)$ defined as
 \begin{eqnarray}
 \mbox{g}(u,s)&=&\int_0^1{\rm d}x~ x\bar{x}\ln{(s-x\bar{x}\bar{u} -i \epsilon)}.
 \end{eqnarray}
We have included the leading electro-weak penguin-type diagrams
induced by the operators $O_1$ and $O_2$ \cite{AC.Ce}
\begin{eqnarray}
 C^\lambda_e=\left[\frac{v_u}{v_t}G^\lambda (s_u)+\frac{v_c}{v_t}
 G^\lambda (s_c)\right]\left(
C_2+\frac{C_1}{N_C}\right).
 \end{eqnarray}

 To calculate the coefficients $a_i$ in Eq. (\ref{coeff}), we have also taken into account the
contributions of the dipole operator $O_{8g}$, which are described
by the functions
\begin{eqnarray}
G_g^0&=&-\int_0^1 {\rm d}u ~\frac{2\Phi_{\parallel}^{V_2}
(u)}{1-u},\nonumber\\
G_g^+&=&-\int_0^1{\rm d}u ~\left( \mbox{g}_{\perp}^{(v){V_2}}(u)+
\frac{\mbox{g}_{\perp}^{\prime(a){V_2}}(u)}{4}\right)\frac{1}{1-u},\nonumber\\
G_g^-&=&\int_0^1 ~\frac{{\rm d}u}{\bar{u}}\left[
-\bar{u}\mbox{g}_{\perp}^{(v){V_2}}(u)+
\frac{\bar{u}\mbox{g}_{\perp}^{\prime(a){V_2}}(u)}{4} +\int_0^u{\rm
d}v\left(\Phi_\parallel^{V_2}(v)-\mbox{g}_\perp^{(v)V_2}(v)\right)
+\frac{\mbox{g}_{\perp}^{(a){V_2}}(u)}{4}\right]\label{o8g},
\end{eqnarray}
here we consider the higher-twist effects $k^{\mu}
=uEn_-^\mu+k_\perp^\mu+\frac{\vec{k}_\perp^2}{4uE}n_+^\mu$ in the
projector of Eq. (\ref{paral}).  The $G_g^-=0$ in Eq. (\ref{o8g}) if
we consider the Wandzura-Wilczek-type relations \cite{WWR}, but we
get $G_g^+\neq 0$ which is different from Ref. \cite{kagan1,ykc}.

With the coefficients in Eq. (\ref{coeff}), we can obtain the decay
amplitudes of the SM  part $\mathcal{A}^{SM}$.   $B\to V V$ decay
amplitudes are given in Appendix A.

\subsection {R-parity violating SUSY effects in the decays}
Flavor changing neutral current (FCNC) is forbidden at tree level in
the SM.  FCNC processes could happen at one loop level, but are
suppressed by Glashow-Iliopoulos-Maiani (GIM) mechanism \cite{GIM}.
New physics effects could be comparable to the SM strength, so that
penguin dominated rare B decays are sensitive to new physics. The
RPV SUSY is an interesting scenario and its possible effects in B
rare nonleptonic decays deserve our studies \cite{report, allanach}.

The R-parity symmetry was first introduced by Farrar and Fayet
\cite{Fayet}, which is assumed  to forbid gauge invariant lepton and
baryon number violating operators \cite{SUSY}. The R-parity of a
particle field is given by $R_p =(-1)^{L+2S+3B}$, where L and B are
lepton and baryon numbers, and S is the spin. However, there is no
deep theoretical motivation for imposing R-parity.  The presence of
RPV could  give very rich phenomenology. Of course, it will get
constraints from its phenomenology. The status of RPV SUSY and
constraints on its parameters could be found in the recent reviews
\cite{report, allanach}.

In the most general superpotential of the minimal supersymmetric
Standard Model (MSSM), the RPV superpotential is given by
\cite{RPVSW}
\begin{eqnarray}
\mathcal{W}_{\spur{R}}&=&\frac{1}{2}\lambda_{[ij]k}
\hat{L}_i\hat{L}_j\hat{E}^c_k+
\lambda'_{ijk}\hat{L}_i\hat{Q}_j\hat{D}^c_k+\frac{1}{2}
\lambda''_{i[jk]}\hat{U}^c_i\hat{D}^c_j\hat{D}^c_k,
\label{rpv}
\end{eqnarray}
where $\hat{L}$ and $\hat{Q}$ are the SU(2)-doublet lepton and quark
superfields and $\hat{E}^c$, $\hat{U}^c$ and $\hat{D}^c$ are the
singlet superfields, while i, j, and k are generation indices and
$c$ denotes a charge conjugate field.

The $\lambda$ and $\lambda'$ couplings in Eq. (\ref{rpv}) break the
lepton number, while the $\lambda''$ couplings break the baryon
number conservation. There are 27 $\lambda'$-type couplings and nine
each of the $\lambda$ and $\lambda''$ couplings as $\lambda_{[ij]k}$
is antisymmetric in i and j, and $\lambda''_{i[jk]}$ is
antisymmetric in j and k. The antisymmetry of the B-violating
couplings, $\lambda''_{i[jk]}$ in the last two indices implies that
there are no operators  generating the $\bar{b}\rightarrow \bar{s}s
\bar{s}$ and $\bar{b}\rightarrow \bar{d} d \bar{d}$ transition.

From Eq. (\ref{rpv}), we can obtain the following four fermion
effective  Hamiltonian due to the exchanging of the sleptons:
\begin{eqnarray}
\mathcal{H}'^{\spur{R}}_{2u-2d}&=&\sum_i\frac{\lambda'_{ijm}
\lambda'^*_{ikl}}{2m^2_{\tilde{e}_{Li}}}
\eta^{-8/\beta_0}(\bar{d}_m\gamma^\mu P_Rd_l)_8(\bar{u}_k
\gamma_\mu P_Lu_j)_8,\nonumber\\
\mathcal{H}'^{\spur{R}}_{4d}&=&\sum_i
\frac{\lambda'_{ijm}\lambda'^*_{ikl}}{2m^2_{\tilde{\nu}_{Li}}}
\eta^{-8/\beta_0}(\bar{d}_m\gamma^\mu P_Rd_l)_8(\bar{d}_k\gamma_\mu
P_Ld_j)_8.
\end{eqnarray}
The four fermion Hamiltonian due to the exchange of the squarks
can be written as
\begin{eqnarray}
\mathcal{H}''^{\spur{R}}_{2u-2d}&=&\sum_n\frac{\lambda''_{ikn}
\lambda''^*_{jln}}{2m^2_{\tilde{d}_{n}}}\eta^{-4/\beta_0}
\left\{\left[(\bar{u}_i\gamma^\mu P_Ru_j)_1(\bar{d}_k\gamma_\mu
P_Rd_l)_1-(\bar{u}_i\gamma^\mu P_Ru_j)_8(\bar{d}_k\gamma_\mu
P_Rd_l)_8\right]\right. \nonumber
\\&&\hspace{3cm}-\left[\left.(\bar{d}_k\gamma^\mu P_Ru_j)_1
(\bar{u}_i\gamma_\mu P_Rd_l)_1
-(\bar{d}_k\gamma^\mu P_Ru_j)_8
(\bar{u}_i\gamma_\mu P_Rd_l)_8\right]\right\},\nonumber\\
\mathcal{H}''^{\spur{R}}_{4d}&=&\sum_n
\frac{\lambda''_{nik}\lambda''^*_{njl}}{4m^2_{\tilde{u}_{n}}}
\eta^{-4/\beta_0}\left[(\bar{d}_i\gamma^\mu
P_Rd_j)_1(\bar{d}_k\gamma_\mu P_Rd_l)_1-(\bar{d}_i\gamma^\mu
P_Rd_j)_8(\bar{d}_k\gamma_\mu P_Rd_l)_8\right].
\end{eqnarray}
Where $P_L=\frac{1-\gamma_5}{2},P_R=\frac{1+\gamma_5}{2},
\eta=\frac{\alpha_s(m_{\hat{f}_i})}{\alpha_s(m_b)}$ and
$\beta_0=11-\frac{2}{3}n_f$. The subscript for the currents
$(j_{\mu})_{1, 8} $ represent  the current in the color singlet and
octet, respectively.  The coefficients $\eta^{-4/\beta_0}$ and
$\eta^{-8/\beta_0}$ are due to the running from the sfermion mass
scale $m_{\hat{f}_i}$ (100 GeV assumed) down to the $m_b$ scale.
Since  it is always assumed in phenomenology for numerical display
that only one sfermion contributes one time, we neglect the mixing
between the operators when we use the renormalization group equation
(RGE) to run $\mathcal{H}^{\spur{R}}$ down to low scale. The
$\mathcal{H}^{\spur{R}}$ for the relevant decay modes are written
down in Appendix B.

Compared  with the operators in the $\mathcal{H}^{SM}_{eff}$, there
are new operators $(\bar{q}_2q_3)_{V\pm A} (\bar{b}q_1)_{V+A}$ in
the $\mathcal{H}^{\spur{R}}$. We will use the $(')$ denote the
$(\bar{q}_2q_3)_{V\pm A} (\bar{b}q_1)_{V+A}$ current contribution.
In the NF approach, the factorizable amplitude can be expressed as
\begin{eqnarray}
  X'^{(BV_1,V_2)}&=&\langle V_2|(\bar{q}_2\gamma_\mu(1-a\gamma_5)q_3)|0\rangle
 \langle V_1|(\bar{b}\gamma^\mu(1+\gamma_5)q_1)|B\rangle.
 \end{eqnarray}

Taking the $V_1$ ($V_2$) meson flying  in the minus (plus)
z-direction and using the sign convention $\epsilon^{0123}=-1$, we
have
\begin{eqnarray}
 X'^{(BV_1,V_2)}=\left \{\begin{array}{ll}-\frac{i
f_{V_2}}{2m_{V_1}}\left[(m^2_{B}-m_{V_1}^2-m_{V_2}^2)(m_B+m_{V_1})
A_1^{BV_1}(m_{V_2}^2)-\frac{2m_{B}^2p_c^2}{m_{B}+m_{V_1}}
A_2^{BV_1}(m_{V_2}^2)\right]\equiv
h'_0,
\\-i
 f_{V_2}m_{V_2}\left[(m_{B}+m_{V_1})A_1 ^{BV_1}(m_{V_2}^2)\pm
\frac{2m_{B}p_c}{m_{B}+m_{V_1}}V ^{BV_1}(m_{V_2}^2)\right]\equiv
h'_{\pm}.
 \end{array}
 \right.
 \end{eqnarray}

Applying the QCDF approach and supposing $V_1$ ($V_2$) is the
recoiled (emitted) meson, we obtain the  vertex corrections
$f^{'\lambda}_I(a)$ and hard spectator scattering corrections
$f^{'\lambda}_{II}(a)$ as follows:
\begin{eqnarray}
 f_I^{'0}(a)&=&12\ln \frac{\mu}{m_b}+18-6(1+a)-\int_0^1{\rm d}u
\Phi_{\parallel}^{V_2} (u)\left(3\frac{1-2u}{1-u}\ln u-3i \pi\right),\nonumber\\
 f_I^{'\pm}(a)&=&12\ln \frac{\mu}{m_b}+18-6(1+a)-\int_0^1{\rm d}u
\left(g_{\perp}^{(v){V_2}} (u)\pm \frac{a
g_{\perp}^{\prime(a){V_2}}
(u)}{4}\right)\left(3\frac{1-2u}{1-u}\ln u-3i \pi\right),\nonumber\\
f^{'0}_{II}(a)&=&\frac{4\pi^2}{N_C}\frac{if_{B}f_{V_1}f_{V_2}}{h^{'}_0}\int_0^1{\rm
d}\xi \frac{\Phi_1^B (\xi)}{\xi}\int_0^1{\rm d}v
\frac{\Phi_{\parallel}^{V_1} (v)}{\bar{v}}\int_0^1{\rm d}u
\frac{\Phi_{\parallel}^{V_2}
(u)}{u},\nonumber\\
f_{II}^{'\pm}(a)&=&-\frac{4
\pi^2}{N_C}\frac{2if_{B}f^{\perp}_{V_1}f_{V_2}m_{V_2}}{m_{B}h^{'}_{\pm}}
(1\pm1)\int_0^1{\rm d}\xi \frac{\Phi_1^B (\xi)}{\xi}\int_0^1{\rm
d}v \frac{\Phi_{\perp}^{V_1}
(v)}{\bar{v}^2}\nonumber\\
&&\times\int_0^1{\rm d}u \left(g_{\perp}^{(v){V_2}}(u)
+\frac{ag_{\perp}^{\prime(a){V_2}} (u)}{4}\right) +\frac{4
\pi^2}{N_C}\frac{2if_{B}f_{V_1}f_{V_2}m_{V_1}m_{V_2}}{m_{B}^2h^{'}_{\pm}}\int_0^1{\rm
d}\xi \frac{\Phi_1^B (\xi)}{\xi}\nonumber\\
&&\times\int_0^1{\rm d}v {\rm d}u\left(g_{\perp}^{(v){V_1}} (v)
\mp\frac{g_{\perp}^{\prime(a){V_1}}
(v)}{4}\right)\left(g_{\perp}^{(v){V_2}} (u)\pm \frac{a
g_{\perp}^{\prime(a){V_2}} (u)}{4}\right)\frac{u+\bar v} {u{\bar
v}^2},
\end{eqnarray}

 Since we are considering the leading
effects of RPV, we need only evaluate the vertex corrections and
the hard-spectator scattering.  The R-parity violating
contribution to the decay amplitudes $\mathcal{A}^{\spur{R}}$  can
be found  in Appendix C.

\subsection{The polarized fraction and branching ratio }
With the QCDF approach, we can get the total decay amplitude
\begin{eqnarray}
\mathcal{A}_\lambda=\mathcal{A}^{SM}_\lambda+\mathcal{A}^{\spur{R}}_\lambda.
\label{amp}
\end{eqnarray}
The expressions for the SM amplitude $\mathcal{A}^{SM}_\lambda $ and
the RPV amplitude $ \mathcal{A}^{\spur{R}}_\lambda$ are presented in
Appendices A and C.  From the amplitude in Eq. (\ref{amp}), the
branching ratio reads
\begin{eqnarray}
\mathcal{B}r(B\rightarrow VV)=\frac{\tau_B |p_c |}{8\pi
m_B^2}(|\mathcal{A}_0 |^2+|\mathcal{A}_+ |^2+|\mathcal{A}_- |^2),
\end{eqnarray}
where $\tau_B$ is the lifetime of B meson, $p_c$ is the center of
mass momentum, and given by
\begin{eqnarray}
|p_c|=\frac{1}{2m_B}\sqrt{[m_B^2-(m_{V_1}+m_{V_2})^2][m_B^2-(m_{V_1}-m_{V_2})^2]}.
\end{eqnarray}
In order to compare the size of helicity amplitudes,
we express the longitudinal polarization fraction
\begin{eqnarray}
\frac{\Gamma_L}{\Gamma}&=&\frac{|\mathcal{A}_0|^2}
{|\mathcal{A}_0|^2+|\mathcal{A}_+|^2+|\mathcal{A}_-|^2}.
\end{eqnarray}
The ratios of $\Gamma_L/ \Gamma$ measure the relative strength  of
longitudinally polarizations  in the decay.

\section{Input Parameters}

{\bf A. Wilson coefficients} \\
To proceed we use the next-to-leading  Wilson coefficients
calculated  in the naive dimensional regularization (NDR) scheme and
at $m_b$ scale
 \cite{coeff}:
\begin{eqnarray}
C_1&=&1.082,~~~C_2=-0.185,~~~C_3=0.014,~~~C_4=-0.035,
~~~C_5=0.009,~~~C_6=-0.041,\nonumber\\
C_7/\alpha_e &=&-0.002,~~~C_8/\alpha_e =0.054,~~~C_9/\alpha_e
=-1.292, ~~~C_{10}/\alpha_e =0.263,~~~C_{8g}=-0.143.\nonumber
\end{eqnarray}

{\bf B. The CKM matrix element} \\
As for the CKM matrix elements, we will use the  Wolfenstein
parametrization \cite{CKM}:
 \begin{equation} \left(\begin{array}{lll}
 V_{ud} & V_{us} & V_{ub}\\
 V_{cd} & V_{cs} & V_{cb}\\
 V_{td} & V_{ts} & V_{tb}\\
 \end{array}\right)
 =\left(\begin{array}{ccc}
 1-\lambda ^2 /2 & \lambda & A\lambda^3(\rho-i \eta)\\
  -\lambda & 1-\lambda^2 /2 & A\lambda^2 \\
 A\lambda^3(1-\rho-i \eta) & -A\lambda^2  & 1\\
 \end{array}\right).
 \end{equation}
We shall use the values \cite{PDG2004}
\begin{eqnarray*}
|V_{cb}|=0.0413,~~~\lambda=0.22,~~~\bar{\rho}=0.20,~~~\bar{\eta}=0.33,
\end{eqnarray*}
where $\bar{\rho}=\rho(1-\frac{\lambda^2}{2})$ and
$\bar{\eta}=\eta(1-\frac{\lambda^2}{2})$.

{\bf C. Masses and lifetime }\\
 For quark masses, which appear in the
penguin loop corrections with regard to the functions $G^\lambda
(s)$, we take
\begin{eqnarray*}
m_u=m_d=m_s=0,~~~~m_c=1.47 ~\mbox{GeV},~~~~m_b=4.8 ~\mbox{GeV}.
\end{eqnarray*}

To compute the branching ratio, the lifetime of $B$ meson and
 masses of meson are also taken from \cite{PDG2004}
\begin{center}
\begin{tabular}{cccc}
$\tau_{B_{u}}=1.671$ ps,& $m_{B_{u}}=5279$ MeV,&$\tau_{B_{d}}=1.536$
 ps,&$m_{B_{d}}=5279$ MeV
,\\
$m_{K^{*\pm}}=892$ MeV,&$m_\phi=1019$ MeV,&$m_{K^{*0}}=896$ MeV,&$m_{\rho}=776$ MeV. \\
\end{tabular}
\end{center}

{\bf D. The LCDAs of the vector meson}\\
 For the LCDAs of the
vector meson, we use the asymptotic form \cite{PV}
\begin{eqnarray}
\Phi_{\parallel}^V (x)&=&\Phi_{\perp}^V
(x)=g_{\perp}^{(a)V}=6x(1-x),\nonumber\\
g_{\perp}^{(v)V} (x)&=&\frac{3}{4}[1+(2x-1)^2].
\end{eqnarray}
As for the two $B$ meson wave functions in Eq. (\ref{projector}), we
consider  only the $\Phi^B_1 (\xi)$  contribution to the
nonfactorizable corrections as done in the literature
\cite{BBNS,BN}, since $\Phi^B_2 (\xi)$ is power suppressed. We adopt
the moments of the $\Phi_1^B (\xi)$ defined in Ref. \cite{BBNS,BN}
for our numerical evaluation:
\begin{equation}
\int_0^1{\rm d}\xi\frac{\Phi_1^B
(\xi)}{\xi}=\frac{m_{B}}{\lambda_B},
\end{equation}
with $\lambda_B=0.46$ GeV \cite{lamdB}. The quantity $\lambda_B$
parameterizes our ignorance about the $B$ meson distribution
amplitudes and thus brings considerable theoretical  uncertainty.

{\bf E. The decay constant and form factors}

The decay constant and form factors are nonperturbative parameters.
They are available from the experimental data or estimated with
theories, such as lattice calculations, QCD sum rules, etc. For the
decay constant, we take the latest light-cone QCD sum rule results
(LCSR) \cite{BallZwicky} in our calculations:
\begin{center}
\begin{tabular}{ccc}
$f_{K^{\ast}}=217$ MeV,&$f_{\rho}=205$ MeV,&$f_{\phi}=231$ MeV
\\
$f^\bot_{K^{\ast}}=156$ MeV,&$f^\bot_{\rho}=147$ MeV,& $
f^\bot_{\phi}=183$ MeV,\\
\end{tabular}
\end{center}
and $f_{B_{u(d)}}=161$ MeV. For the form factors involving the $B\to
K^{\ast}$ and $B\to\rho$ transition, we adopt the results given by
\cite{BallZwicky}
 \begin{eqnarray*}
  A_1^{B_{u(d)}K^{\ast}}(0)&=&0.292,~
 A_2^{B_{u(d)}K^{\ast}}(0)=0.259,~
 V^{B_{u(d)}K^{\ast}} (0) =0.411,\nonumber\\
 A_1^{B_{u(d)}\rho}(0)&=&0.242,~~
 A_2^{B_{u(d)}\rho}(0)=0.221,~~
 V^{B_{u(d)}\rho} (0) =0.323.
 \end{eqnarray*}

\section{Numerical results and Analysis}
In this section we will give our estimations in the SM and compare
with the relevant experimental data, then we show the RPV  effects
to the branching ratios and longitudinal polarization fractions.
With the RPV effects we can do two things. First, it is possible to
have small longitudinal polarization fractions for the pure penguin
processes completely different from the SM predictions. Second, one
can obtain more stringent bounds on the product combinations of RPV
couplings if the measured values of the decay modes are consistent
with the SM predictions.

In the numerical analysis, we assume that only one sfermion
contributes one  time with a mass of 100 GeV.  So for other values
of the sfermion masses, the bounds on the couplings in this paper
can be easily obtained by  scaling them by factor
$\tilde{f}^2\equiv (\frac{m_{\tilde{f}}}{100GeV})^2$. We consider
all possible sfermion contributions when we obtain the bounds on
the couplings.

\subsection{ $B^0 \to \phi K^{*0}$ and $B^+ \to \phi K^{*+}$}

$B^0 \to \phi K^{*0}$ and   $B^+ \to \phi K^{*+}$ decays are
induced by the underlying quark transition  $\bar{b}\rightarrow
\bar{s}s\bar{s}$  at one loop level. In the SM, these decays are
 called pure penguin processes. However, they could happen at tree level
 in SUSY models with RPV.    Our estimations in the SM and
recent data from the \textit{BABAR} and Belle collaborations are
presented
in the Table I and II.\\

\begin{table}[ht]
\centerline{\parbox{16cm}{{Table I: Branching ratios of $B\to \phi
K^*$ decay modes.}}} \vspace{0.1cm}
\begin{center}
\begin{tabular}{ccccc}\hline\hline
Mode& \textit{BABAR}$(\times 10^6)$ & Belle$(\times 10^6)$ &
Average$(\times 10^6)$& QCDF in SM$(\times 10^6)$\\ \hline
  $B^+_u\rightarrow \phi K^{*+}$& $12.7^{+2.2}_{-2.0}+1.1
   ~\cite{0307026}$&$6.7^{+2.1+0.7}_{-1.9-1.0} ~\cite{0307014}$&$9.5\pm1.7$&4.60\\
  $B^0_d\rightarrow \phi K^{*0}$& $9.2\pm0.9\pm0.5
   ~\cite{0408017}$&$10.0^{+1.6+0.7}_{-1.5-0.8} ~\cite{0307014}$&$9.4\pm0.9$&4.21\\
  \hline
 \end{tabular}
\end{center}
\centerline{\parbox{16cm}{Table II: The longitudinal  polarization
fractions of $B\to \phi K^*$ decay modes.}}
\begin{center}
 \begin{tabular}{ccccc}\hline\hline
Mode& \textit{BABAR} & Belle & Average& QCDF in SM\\ \hline
  $B^+_u\rightarrow \phi K^{*+}$& $0.46\pm0.12\pm0.03  ~\cite{0307026}$&
  $0.51\pm0.08\pm0.03 ~\cite{0503013}$&$0.49\pm0.07$&0.861\\
  $B^0_d\rightarrow \phi K^{*0}$& $0.52\pm0.05\pm0.02
   ~\cite{0408017}$&$0.45\pm0.05\pm0.02 ~\cite{0503013}$&$0.49\pm0.04$&0.861\\
  \hline
 \end{tabular}
\end{center}
\end{table}

\begin{figure}[h]
\begin{center}
\begin{tabular}{cc}
\includegraphics[scale=0.81]{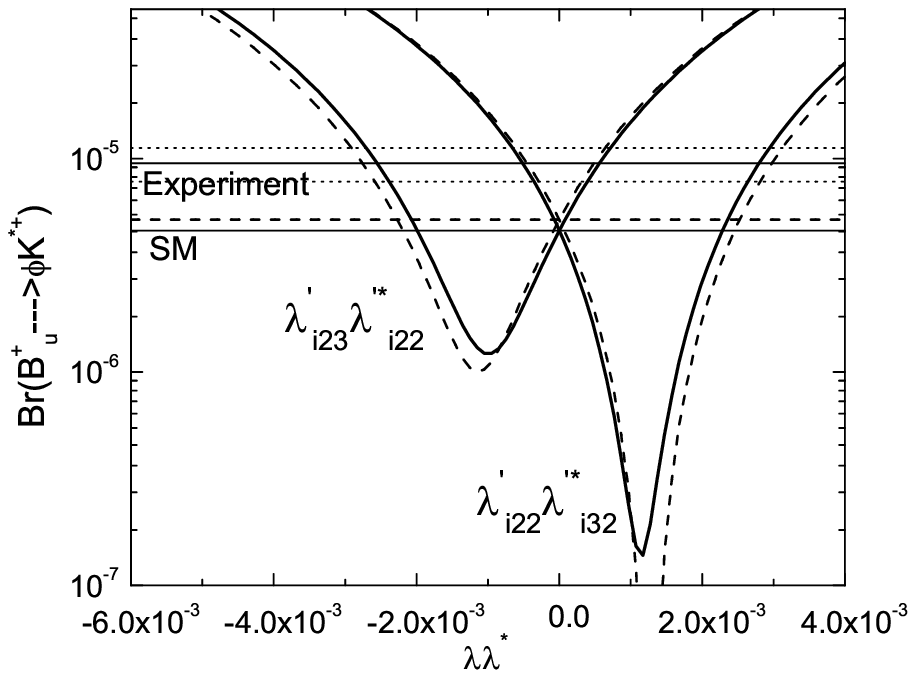}&
\includegraphics[scale=0.81]{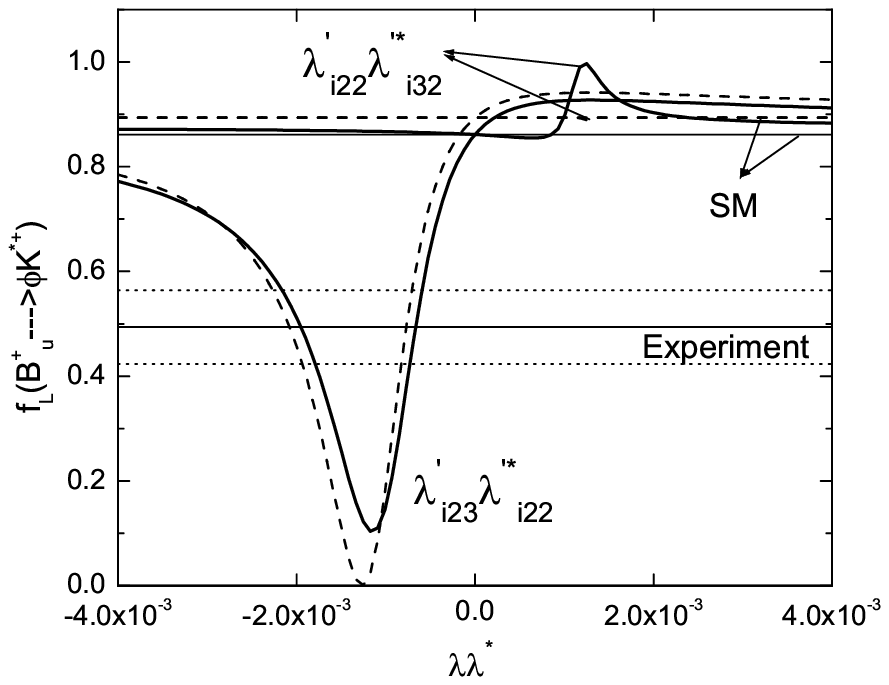}\\
\includegraphics[scale=0.81]{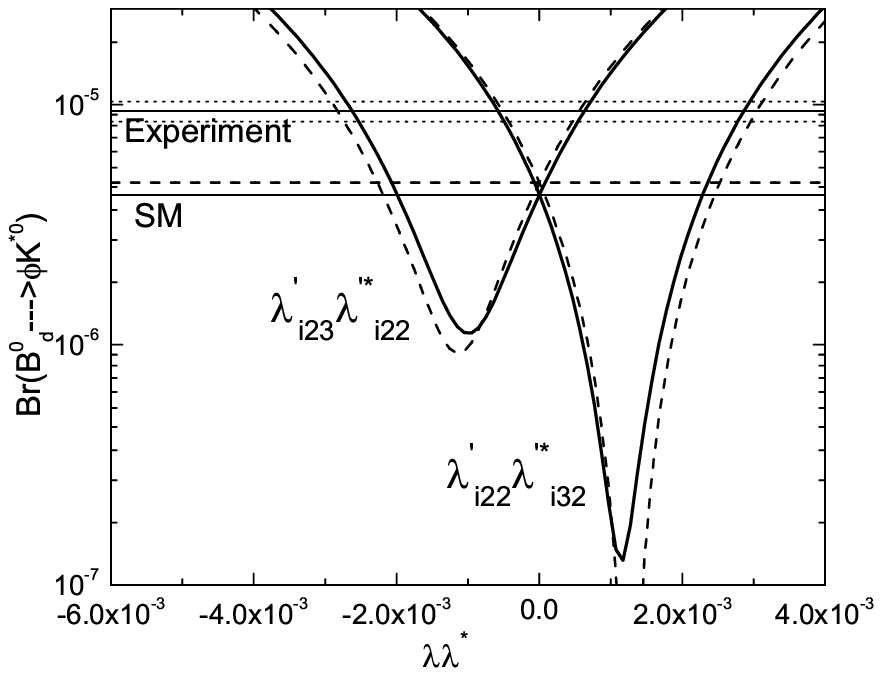}&
\includegraphics[scale=0.81]{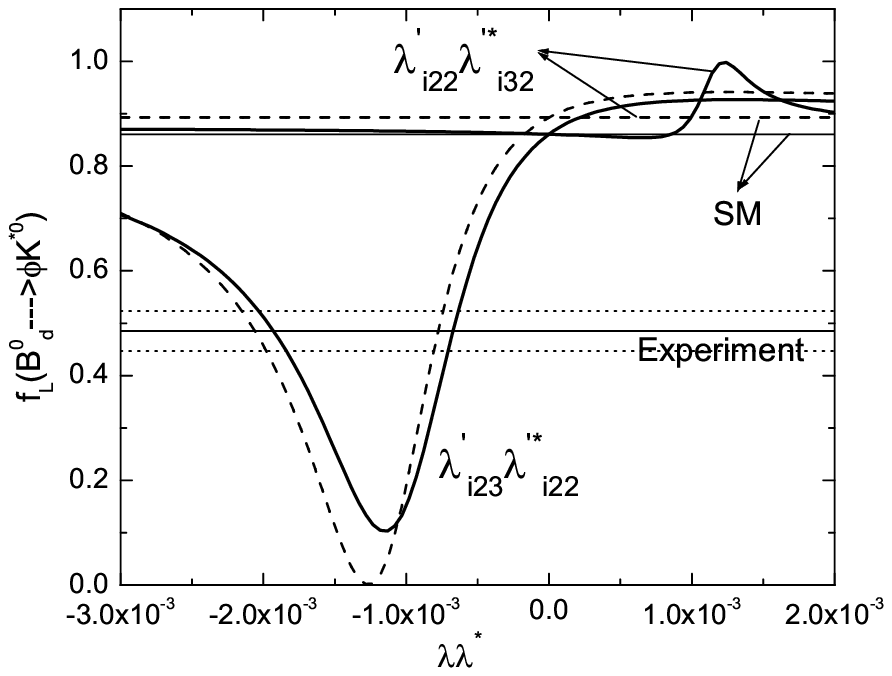}
\end{tabular}
\end{center}
\caption{\small The branching ratios and the longitudinal
polarization fractions for $B\to\phi K^{*}$ as  functions of the
RPV couplings $\lambda'\lambda'^*$. The solid curves represent our
theoretical results by QCDF. The dash lines are the predictions by
NF for comparisons. The horizontal solid lines are the
experimental data and the SM predictions with QCDF as labelled
respectively. The horizontal dot lines represent the $1\sigma$
error-bar of the measurements.(The same in Fig.2 and 3).}
\end{figure}

From Table.I and II, we can see that the predictions from the two
factorization framework are very similar. It is also noted that the
latest LCSR \cite{BallZwicky} results for  factors are smaller than
their pervious values \cite{ball}.  With the form factors,   we find
that the SM predictions for the branching ratios are smaller that
the \textit{BABAR} \cite{0307026, 0408017} and Belle
\cite{0307014,0503013} measurements, and  the longitudinal
polarizations are as large as $\sim 90\%$ in contrast to $\sim 50\%$
measured by \textit{BABAR} and Belle.

Now we turn to RPV effects which may give a possible solution to the
polarization anomaly. Using the formula derived in previous
sections, we can calculate RPV effects. The RPV effects to the
branching ratios and longitudinal polarization fractions of the
$B\to \phi K^*$ decays are  displayed by curves in the Fig.1.  We
find that the $\lambda'_{i22}\lambda'^*_{i32}$ term could not
enhance transverse polarization because the corresponding current
has a structure $\bar{s}\gamma_{\mu}(1+\gamma_5 )s
\bar{b}\gamma^{\mu}(1-\gamma_5)s$. Its matrix  element is the same
as the SM one, since $\langle \phi| \bar{s}\gamma_{\mu}\gamma_5 s
|0\rangle=0$. However, the $\lambda'_{i23}\lambda'^*_{i22}$ term
could provide a solution to the polarization anomaly because its
current has the structure $\bar{s}\gamma_{\mu}(1-\gamma_5 )s
\bar{b}\gamma^{\mu}(1+\gamma_5)s$. The right handed $(\bar{b}s)$
current will flip the signs of the axial parts of the matrix
$\langle K^* |\bar{b}\gamma^{\mu}(1-\gamma_5)s|B\rangle$ appearing
in the SM contributions. We also find that
$|\lambda'_{i23}\lambda'^*_{i22}|< 3.0\times
10^{-3}\tilde{\nu}^2_{Li}$ by the branching ratios; however, there
is a much stronger bound from the polarization ratios.

 Combining both the branching ratios and
the polarization ratios which we get by QCDF (the same as follow),
the resolution is obtained in a very narrow parameter interval
$|\lambda'_{i23}\lambda'^*_{i22}|\in
[1.5\times10^{-3}\tilde{\nu}^2_{Li},2.1\times
10^{-3}\tilde{\nu}^2_{Li}]$.
  Fortunately, the parameter space is
not ruled out yet, the existing upper limit is
$|\lambda'_{i23}\lambda'^*_{i22}|< 2.3\times 10^{-3}$  \cite{report,
RPVstudy}. We note that similar strength of the RPV couplings  is
also needed in the recent studies to solve the CP problem in $B\to
\phi K$ \cite{datta} and the $\eta\prime$ puzzle in $B\to \eta\prime
K$ \cite{cskim}.

\subsection {$B^+_u\to \rho K^{*}$ decays }

 $B^+_u\to \rho K^*$ decays are due to  $\bar{b}\rightarrow \bar{u}u\bar{s}$
or $\bar{b}\rightarrow \bar{d}d\bar{s}$ transitions   at the quark
level. The $B^+_u \to \rho^+ K^{*0}$ decay is a pure penguin process
in the SM. Its longitudinal polarization fraction were measured to
be unexpected low $\sim 0.5$ very recently by Belle \cite{Belle BJ},
which  is inconsistent with the SM prediction. While the $B^+_u \to
\rho^0 K^{*+}$ has both tree and penguin amplitude, experimental
measurements by \textit{BABAR} \cite{0307026}  have shown the decay
dominated by longitudinal polarization, which is consistent with the
SM predictions.

For comparison, our estimations in the SM and recent data from
experiments are presented in the Table III and IV.\\
\begin{table}[ht]
\centerline{\parbox{14cm}{Table III: Branching ratios of $B\to
\rho K^*$ decay modes.}} \vspace{0.1cm}
\begin{center}
 \begin{tabular}{cccc}\hline\hline
Mode& \textit{BABAR} $(\times 10^6)$ & Belle$(\times 10^6)$ & QCDF
in SM$(\times 10^6)$\\ \hline
 $B^+_u\rightarrow \rho^0 K^{*+}$& $10.6^{+3.0}_{-2.6}\pm2.4~\cite{0307026}$&$\cdots$&2.58\\
 $B^+_u\rightarrow \rho^+ K^{*0}$& $\cdots$&$6.6\pm2.2\pm0.8 ~\cite{Belle BJ}$&3.79\\
  \hline
 \end{tabular}
\end{center}
\end{table}
\begin{table}[ht]
\centerline{\parbox{14cm}{Table IV: The longitudinal polarization
fractions of $B\to \rho K^*$ decay modes.}} \vspace{0.1in}
\begin{center}
 \begin{tabular}{ccccc}\hline\hline
Mode& \textit{BABAR} & Belle &  QCDF in SM\\ \hline
   $B^+_u\rightarrow \rho^0 K^{*+}$& $0.96^{+0.04}_{-0.15}\pm0.04 ~\cite{0307026}$
   &$\cdots$&0.906\\
 $B^+_u\rightarrow \rho^+ K^{*0}$& $\cdots$ &$0.50\pm0.19^{+0.05}_{-0.07}
 ~\cite{Belle BJ}$&0.905\\
  \hline
 \end{tabular}
\end{center}
\end{table}

From Table III, we can see the SM predictions for the branching
ratios are smaller than the measurement by \textit{BABAR} and Belle.
As shown in Table IV,  for the longitudinal  polarization fraction
in the tree dominant decay $B^+ \to \rho^0 K^{*+}$,
 the SM prediction agrees with the \textit{BABAR} measurement very well.
However, for the pure penguin process $B^+ \to \rho^+ K^{*0}$, the
SM predicts dominant  longitudinal polarization, which is in
contrast to Belle measurement \cite{Belle BJ}.

In the $B^+_u\rightarrow \rho^0 K^{*+}$ decay,  since the quark
content of $\rho^0$ is $(u\bar{u}-d\bar{d})/\sqrt{2}$, the decay
could be induced by superpartners of  both up-type and down-type
fermions. For example, $\bar{b}\rightarrow \bar{d}d\bar{s}$ could
be induced by sneutrino, while $\bar{b}\rightarrow
\bar{u}u\bar{s}$ could be induced by slepton with the same
$\lambda'_{i13}\lambda'^*_{i12}$ product. We take
$\frac{\lambda'_{i13}\lambda'^*_{i12}}{m^2_{\tilde{e}_i}}$ and
$\frac{\lambda'_{i13}\lambda'^*_{i12}}{m^2_{\tilde{\nu}_i}}$
contribute to $\bar{b}\rightarrow \bar{u}u\bar{s}$ and
$\bar{b}\rightarrow \bar{d}d\bar{s}$ at the same time, so the
effects of $\lambda'_{i13}\lambda'^*_{i12}$ will be eliminated if
taking $m_{\tilde{e}_i}=m_{\tilde{\nu}_i}$.
\begin{figure}[t]
\begin{center}
\begin{tabular}{cc}
\includegraphics[scale=0.81]{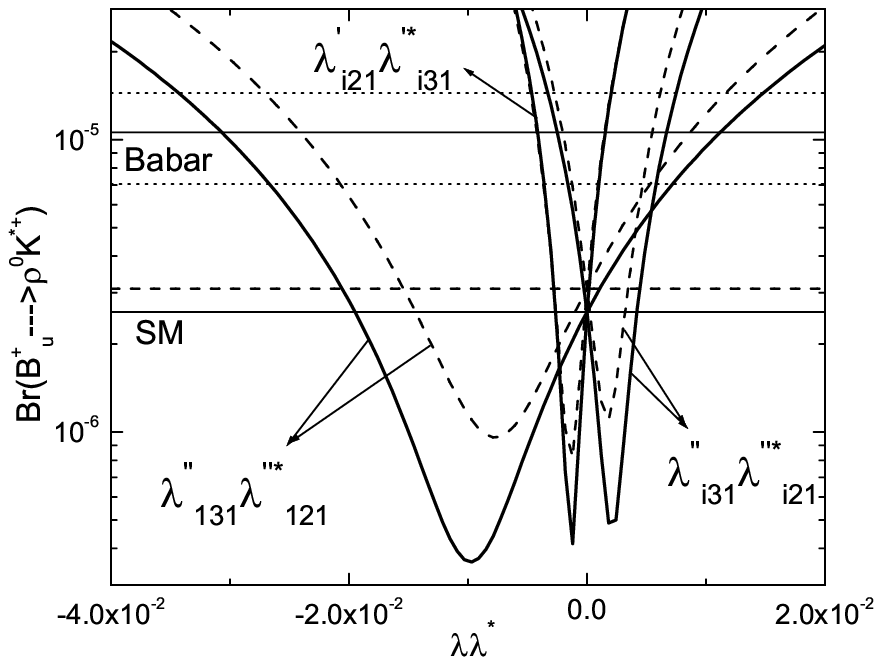}&
\includegraphics[scale=0.81]{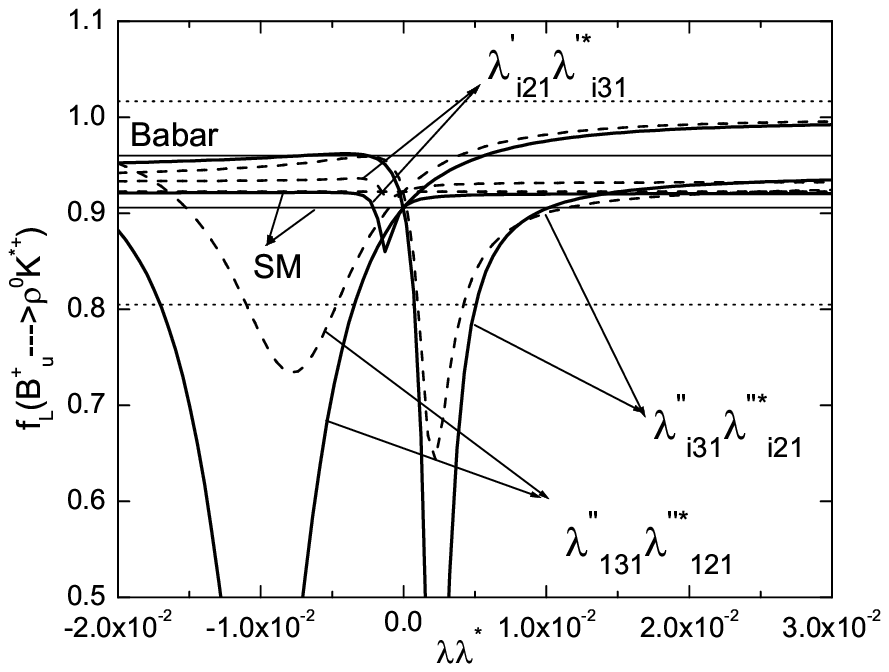}\\
\includegraphics[scale=0.81]{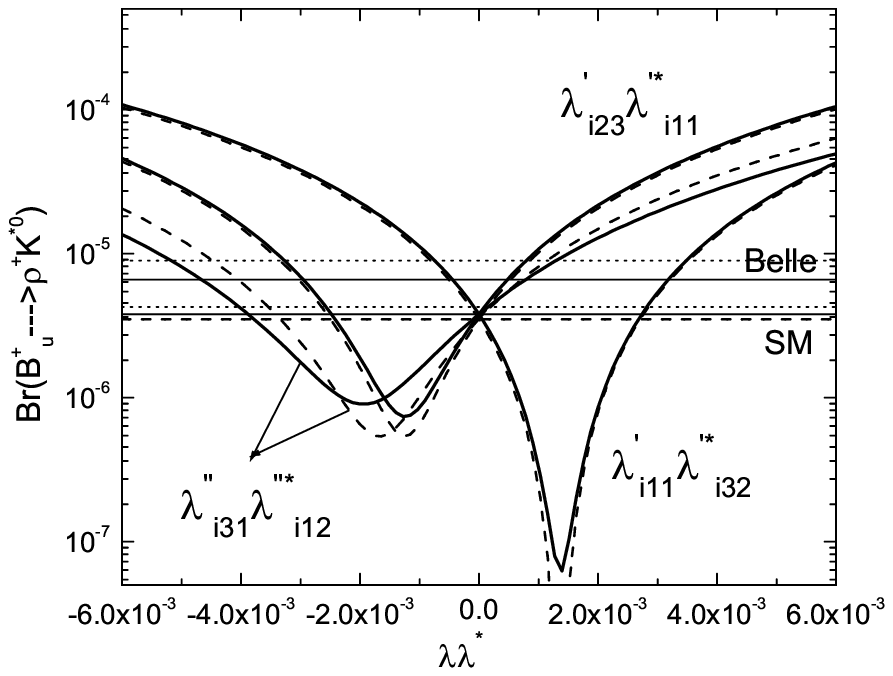}&
\includegraphics[scale=0.81]{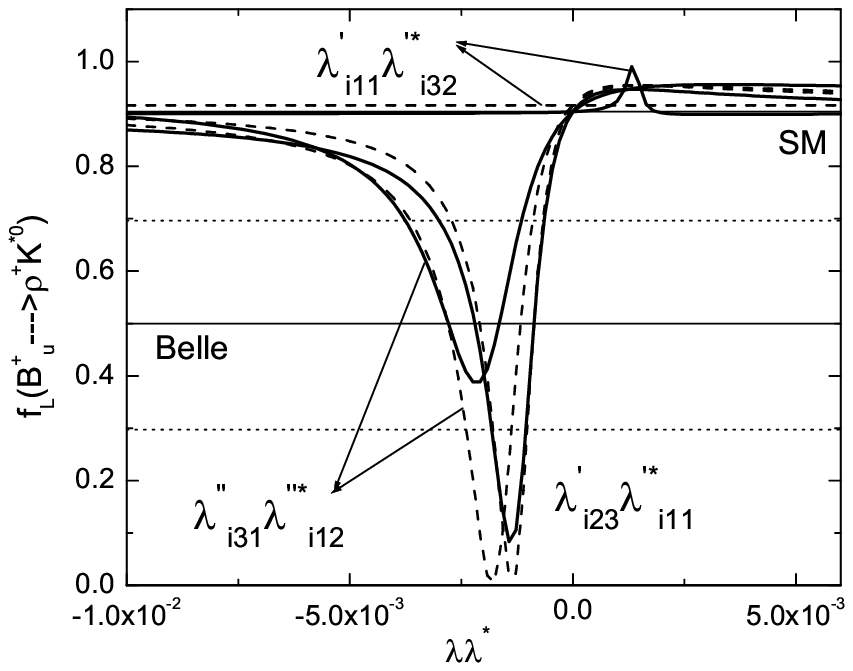}
\end{tabular}
\end{center}
\caption{The branching ratios and longitudinal polarization
fractions for $B\rightarrow \rho K^{*}$ as functions of the RPV
couplings $\lambda'\lambda'^*$ and $\lambda''\lambda''^*$.}
\end{figure}

Our results for the RPV contributions to $B^+_u\rightarrow \rho
K^{*}$ are summarized in Fig.2. We can get a lot of information from
this figure. We find that the longitudinal polarization in $B^+_u
\to \rho^0 K^{*+}$ is sensitive to
$|\lambda''_{131}\lambda''^*_{121}|$ and
$|\lambda''_{i31}\lambda''^*_{i21}|$ arising from exchanging
$\tilde{d}$ and $\tilde{u}_i$ between the quark currents
$\bar{u}\gamma_{\mu}(1+\gamma_5 )u \otimes
\bar{b}\gamma^{\mu}(1+\gamma_5)s$ and
$\bar{d}\gamma_{\mu}(1+\gamma_5 )d\otimes
\bar{b}\gamma^{\mu}(1+\gamma_5)s$, respectively, and insensitive to
the effect from exchanging $\tilde{\nu}_i $ between the quark
currents $\bar{d}\gamma_{\mu}(1+\gamma_5 )d \otimes
\bar{b}\gamma^{\mu}(1-\gamma_5)s$.
 However, the pure penguin process $B^+_u\rightarrow \rho^+
K^{*0}$ is sensitive to the RPV couplings
$|\lambda''_{i31}\lambda''^*_{i12}|$ and
$|\lambda'_{i23}\lambda'^*_{i11}|$  which associate with operators
 $\bar{d}\gamma_{\mu}(1+\gamma_5 )s \otimes
\bar{b}\gamma^{\mu}(1+\gamma_5)d$ and
$\bar{d}\gamma_{\mu}(1-\gamma_5 )s \otimes
\bar{b}\gamma^{\mu}(1+\gamma_5)d$,  respectively.

From Fig.2, the polarization problem in $B^+_u\to \rho^0 K^{*+}$
could be solved by RPV effects. However,
 combining the constraints from  polarization fraction and branching
 ratios of $B^+_u\to \rho K^{*}$ ,
 we obtain very narrow parameters spaces for relevant coupling constants.
 It implies that these couplings
could be pinned down or ruled out by refined measurements in the
very near future at \textit{BABAR} and Belle. Our constraints are
listed in Table V.
\begin{table}[t]
\centerline{\parbox{17cm}{{Table V: Bounds on the quadric coupling
constant products. For comparison, the exiting bounds are listed
in last column.}}} \vspace{0.1cm}
\begin{center}
\begin{tabular}{cllcc}\hline
Couplings&~~~~~~~~~~~~~~~~~~~~~~~~~~Bounds& &Process& Previous bounds\\
\hline $|\lambda'_{i21}\lambda'^*_{i31}|$&$^{\leq 5.1 \times
10^{-3}\tilde{\nu}^2_{Li}}_{\leq 2.5 \times
10^{-3}\tilde{\nu}^2_{Li}}$ &$ ^{{\rm for}~~
\lambda'_{i21}\lambda'^*_{i31}<0}_{{\rm for}~~
\lambda'_{i21}\lambda'^*_{i31}>0}$&$B^+_u\rightarrow \rho^0
K^{*+}$&$8.2\times 10^{-4}$~\cite{allanach}
\\\hline
$|\lambda''_{i31}\lambda''^*_{i21}|$&$_{[4.0\times
10^{-3}\tilde{u}_i^2, 8.3\times 10^{-3}\tilde{u}_i^2] {\rm and} \leq
1.1\times 10^{-3}\tilde{u}_i^2}^{\leq 4.1 \times
10^{-3}\tilde{u}_i^2}$& $_{{\rm
for}~~\lambda''_{i31}\lambda''^*_{i21}>0}^{{\rm
for}~~\lambda''_{i31}\lambda''^*_{i21}<0}$& $B^+_u\rightarrow \rho^0
K^{*+}$&$1.0\times 10^{-2}$~\cite{hexg}
\\\hline
$|\lambda''_{131}\lambda''^*_{121}|$
&$^{\leq1.8\times10^{-2}\tilde{d}^2}_{[ 1.6\times
10^{-2}\tilde{d}^2, 3.8\times 10^{-2}\tilde{d}^2]{\rm and} \leq
4.7\times10^{-3}\tilde{d}^2}$&$^{{\rm for}
~~\lambda''_{131}\lambda''^*_{121}>0}_{{\rm for}
~~\lambda''_{131}\lambda''^*_{121}<0}$ & $B^+_u\rightarrow \rho^0
K^{*+}$&$2\times 10^{-8}$~\cite{allanach}
\\\hline \hline
$|\lambda'_{i23}\lambda'^*_{i11}|$&$^{\leq1.5\times
10^{-3}\tilde{\nu}^2_{Li}}_{[1.6\times
10^{-3}\tilde{\nu}^2_{Li},4.0\times 10^{-3}\tilde{\nu}^2_{Li}] {\rm
and} \leq 1.2\times 10^{-3}\tilde{\nu}^2_{Li} }$&$_{{\rm for
}~~\lambda'_{i23}\lambda'^*_{i11}<0}^{{\rm for
}~~\lambda'_{i23}\lambda'^*_{i11}>0}$ &$B^+_u\rightarrow \rho^+
K^{*0}$&$2.1\times 10^{-3}$~\cite{hexg}
\\\hline
$|\lambda'_{i11}\lambda'^*_{i32}|$&$^{\leq 1.4 \times
10^{-3}\tilde{\nu}^2_{Li}}_{\leq 4.1 \times
10^{-3}\tilde{\nu}^2_{Li}}$&$_{{\rm
for}~~\lambda'_{i11}\lambda'^*_{i32}>0}^{{\rm
for}~~\lambda'_{i11}\lambda'^*_{i32}<0}$ & $B^+_u\rightarrow \rho^+
K^{*0}$&$4.7\times 10^{-4}$~\cite{deltaK}
\\\hline
$|\lambda''_{i31}\lambda''^*_{i21}|$&$^{\leq 2.5\times
10^{-3}\tilde{u}_i^2}_{\leq 6.3\times 10^{-3}\tilde{u}_i^2}$&$_{{\rm
for}~~ \lambda''_{i31}\lambda''^*_{i21}>0}^{{\rm for}~~
\lambda''_{i31}\lambda''^*_{i21}<0}$&$B^+_u\rightarrow \rho^+
K^{*0}$&$1.0\times 10^{-2}$~\cite{hexg}
\\\hline
\end{tabular}
\end{center}
\end{table}

For comparison, we have also listed existing bounds on these quadric
coupling constant products. We can see that the product
$|\lambda''_{131}\lambda''^*_{121}|$ is severely constrained by
double nucleon decay \cite{allanach}. Our bounds on
$|\lambda'_{i11}\lambda'^*_{i32}|$ is weaker than that by $\Delta
m_K$ \cite{deltaK}. These couplings also contribute to $B\to K\pi$
decays. In an interesting study \cite{hexg}, constraints on these
couplings have been derived from the experiment measurement of
branching ratios of $B\to K\pi$. As shown in Table V, our bounds are
lower than theirs \cite{hexg}. However, in  view of new data from
\textit{BABAR} and Belle, it would be interesting to investigate
whether the RPV couplings could solve the known $\pi K$ puzzle
\cite{piki, piki2}.

\subsection {$B\to \rho \rho$ decays}

Recent measurements of tree dominated vector-vector charmless modes
$B^+_u\to \rho^0 \rho^+$ and $B^0_d\to \rho^+ \rho^-$ by
\textit{BABAR}
 \cite{0307026, 0404029} and Belle \cite{0306007} have shown that the
decays are dominated by longitudinal polarizations, which are just
as the SM expectations.

Our estimations in the SM   and recent data from the \textit{BABAR}
and Belle collaborations are presented in  Table VI and VII. We find
that the SM predictions for both  branching ratios and longitudinal
polarization fractions agree with the \textit{BABAR} and Belle
measurements very well. Therefore, these decays could give strong
constraints on relevant RPV couplings.

\begin{table}[t]
\centerline{\parbox{16cm}{Table VI: Branching ratios of $B\to \rho
\rho$ decay modes.}}
\begin{center}
 \begin{tabular}{ccccc}\hline\hline
Mode& \textit{BABAR}$(\times 10^6)$ & Belle$(\times 10^6)$ &
Average$(\times 10^6)$& QCDF in SM$(\times 10^6)$\\ \hline
   $B^+_u\rightarrow \rho^0 \rho^+$& $22.5^{+5.7}_{-5.4}
   \pm5.8~\cite{0307026}$&$31.7\pm7.1^{+3.8}_{-6.7}~\cite{0306007}$&$26.2\pm6.2$&15.30\\
 $B^0_d\rightarrow \rho^+ \rho^-$& $30\pm4\pm5~\cite{0404029}$&$\cdots$&&25.75\\
 \hline
 \end{tabular}
\end{center}
\centerline{\parbox{16cm}{Table VII: The longitudinal polarization
fractions of $B\to \rho \rho$ decay modes.}}
\begin{center}
 \begin{tabular}{ccccc}\hline\hline
Mode& \textit{BABAR} & Belle & Average& QCDF in SM\\ \hline
  $B^+_u\rightarrow \rho^0 \rho^+$& $0.97^{+0.03}_{-0.07}
  \pm0.04~\cite{0307026}$&$0.948\pm0.106\pm0.021~\cite{0306007}$&$0.964\pm0.056$&0.941\\
 $B^0_d\rightarrow \rho^+ \rho^-$& $0.99\pm0.03^{+0.04}_{-0.03}
 ~\cite{0404029}$&$\cdots$&&0.935\\
 \hline
 \end{tabular}
\end{center}
\end{table}

\begin{figure}[ht]
\begin{center}
\begin{tabular}{cc}
\includegraphics[scale=0.81]{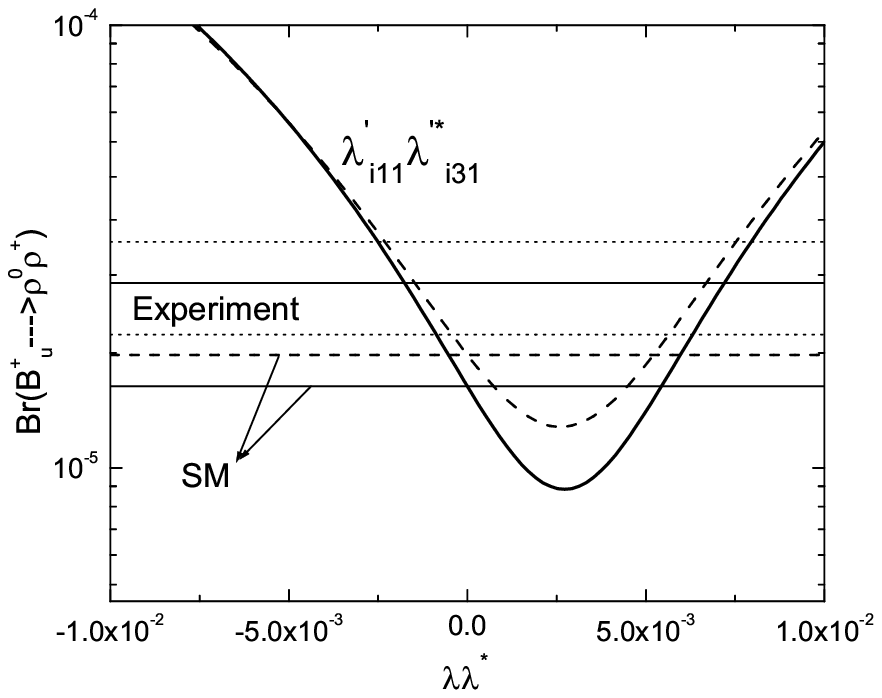}&
\includegraphics[scale=0.81]{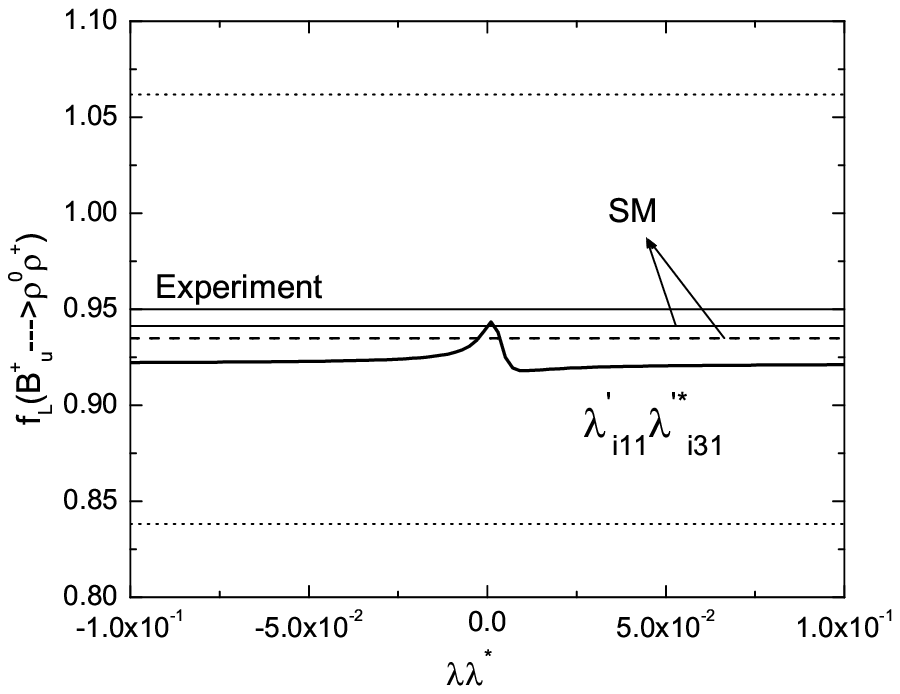}\\
\includegraphics[scale=0.81]{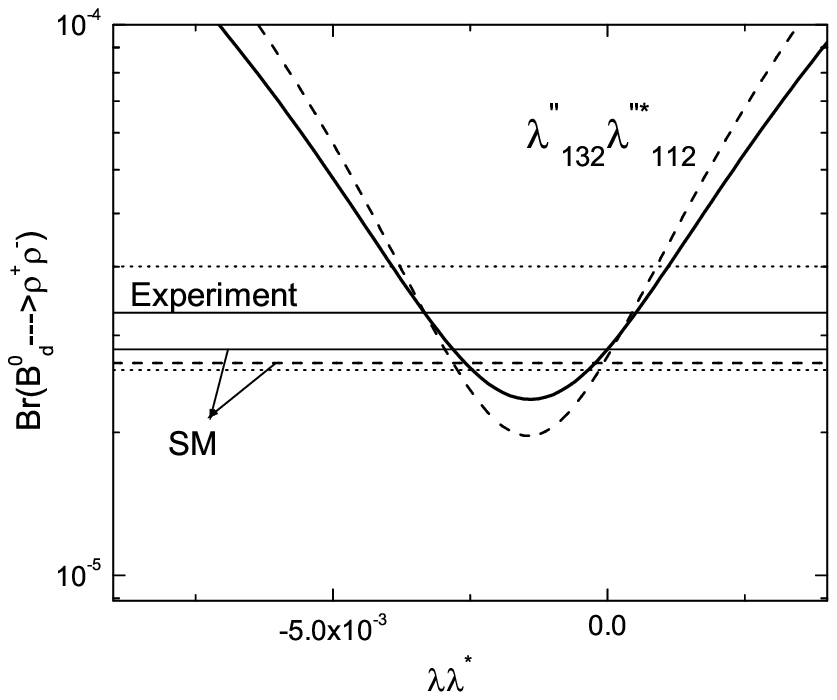}&
\includegraphics[scale=0.81]{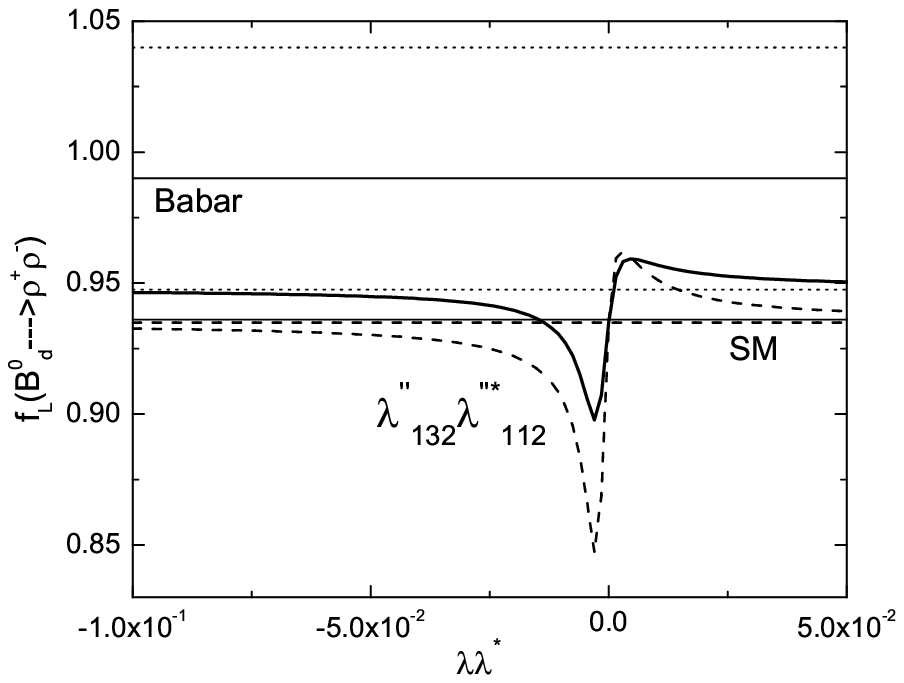}
\end{tabular}
\end{center}
\caption{The branching ratios and longitudinal polarizations  for
$B\rightarrow \rho \rho$ as a function of the RPV couplings
$\lambda'\lambda'^*$ and $\lambda''\lambda''^*$.}
\end{figure}

The RPV contributions  are presented in Fig.3. From the figure, we
can know that longitudinal polarization fractions for $B\rightarrow
\rho \rho$ are  not  sensitive to the RPV couplings since they are
tree dominated decay in the SM.  By the decay $B^+_u\rightarrow
\rho^0 \rho^+$, we get the bounds on
$|\lambda'_{i11}\lambda'^*_{i31}|$:
\begin{eqnarray}
 |\lambda'_{i11}\lambda'^*_{i31}|\in \hspace{1cm} ^{[6.3\times
10^{-3}\tilde{\nu}^2_{Li},~ 8.0\times10^{-3}\tilde{\nu}^2_{Li}]
\hspace{0.5cm} {\rm for\ }
\lambda'_{i11}\lambda'^*_{i31}>0}_{[8.6\times10^{-4}\tilde{\nu}^2_{Li},~
2.5\times10^{-3}\tilde{\nu}^2_{Li}] \hspace{0.5cm}{\rm for\ }
\lambda'_{i11}\lambda'^*_{i31}<0} \hspace{1cm} {\rm by
}~~B^+_u\rightarrow \rho^0 \rho^+.
\end{eqnarray}
The existing bounds on the products by $B\to \pi \pi$ decays are
$|\lambda'_{i11}\lambda'^*_{i31}|<1.6\times 10^{-2}$ \cite{hexg}.
The parameter spaces constrained by $B^+_u\rightarrow \rho^0 \rho^+$
are very narrow, which could be closed easily be future refined
measurements if the tight bounds from branching ratio and
polarization ratio do not overlap.  This could not be done by the
$B\to PP, PV$ decays. We see again the rich phenomena of $B\to VV$
decays and its power for bounding new physics.

We note that RPV coupling $\lambda''_{132}\lambda''^*_{112}$ is
eliminated because the factorizable amplitudes
$X'^{(B^+_u\rho^0,\rho^+)}=X'^{(B^+_u\rho^+,\rho^0)}$ and
cancelled each other.  The effect of
$\lambda'_{i13}\lambda'^*_{i11}$ also is eliminated for the same
reason in $B^+_u\to \rho^0K^{*+}$.

The decay $B^0_d\rightarrow \rho^+ \rho^-$ give  constraint
  $\lambda''_{132}\lambda''^*_{112}\in[9.3
\times 10^{-4}\tilde{s}^2, 1.1\times10^{-3}\tilde{s}^2]$. However,
this parameter space is already ruled out in  previous studies
\cite{report,allanach}. Again, we see the standard model works well
for tree dominant B decays.

Through  the numerical analysis in this section, we can conclude
that the experimental value of the  polarization anomaly in pure
penguin decays can be solved  with RPV effects; however, the
solution parameter spaces are always very narrow.

\section{Conclusions}

Motivated by the polarization anomaly observed recently by
\textit{BABAR} and Belle, we have studied $B \rightarrow VV $ modes
with QCD factorization approach both in the SM and in RPV SUSY
theories. We have found a set of RPV couplings can give possible
solution to the polarization anomaly. However, the windows of the
RPV couplings intervals  are found to be always very narrow. It
implies that these couplings might be pinned down from  the rich
polarization phenomena in these decays. However, it  also implies
the window could be closed easily with refined measurements from
\textit{BABAR} and Belle in the near future.

Since the hadronic dynamics for B nonleptonic decays are generally
tangled with nonperturbative dynamics, we need to known how to
separate perturbative and nonperturbative dynamics; of course, we
also need to know the value for the parameters of nonperturbative
dynamics to give reliable predictions based on electroweak theories.
At the present stage, QCDF is a working scheme. Generally, we can
believe that QCDF calculations for polarization fractions could be
much more accurate than that for branching ratios, since many
uncertainties could be cancelled in the fractions. Therefore the
constraints from polarization measurements would be more
well-founded than those from branching ratio measurements.

Comparing our prediction with the recent experimental data,
 we have obtained bounds on the relevant
products of  RPV couplings.  We find that many
 bounds  are stronger than the existing limits
 \cite{report,RPVstudy}, which may be useful for further studying
the RPV SUSY.

In conclusion, we have shown that RPV SUSY could give possible
solution to the polarization anomaly in pure penguin decays $B\to
\phi K^*$ and $B^+ \to \rho^+ K^{*0}$ observed by \textit{BABAR} and
Belle.

\section*{Acknowledgments}

We thank Alex  Kagan for helpful communications and comments on
this paper.   The work is supported in part by National Science
Foundation under contract No.10305003, Henan Provincial Foundation
for Prominent Young Scientists under contract No.0312001700 and in
part by the project sponsored by SRF for ROCS, SEM.\\

 \begin{appendix}
 \begin{center}
 {\LARGE{\bf Appendix}}
 \end{center}
 \section{\hspace{-0.6cm}. The amplitudes in the SM}\vspace{-1cm}
\begin{eqnarray}
\mathcal{A}^{SM}_\lambda(B^+_u\rightarrow \phi
K^{*+})&=&-\frac{G_F}{\sqrt{2}}V^*_{tb}V_{ts}[a^\lambda_3+a^\lambda_4
+a^\lambda_5-\frac{1}{2}(a^\lambda_7+a^\lambda_9+a^\lambda_{10})]X^{(B^+_uK^{*+},
\phi)},\\
\mathcal{A}^{SM}_\lambda(B^0_d\rightarrow \phi
K^{*0})&=&-\frac{G_F}{\sqrt{2}}V^*_{tb}V_{ts}[a^\lambda_3+a^\lambda_4
+a^\lambda_5-\frac{1}{2}(a^\lambda_7+a^\lambda_9+a^\lambda_{10})]X^{(B^0_dK^{*0},
\phi)},\\
\mathcal{A}^{SM}_\lambda(B^+_u\rightarrow \rho^0
K^{*+})&=&\frac{G_F}{2}\left\{[V^*_{ub}V_{us}a^\lambda_2-V^*_{tb}V_{ts}
\frac{3}{2}(a^\lambda_7+a^\lambda_9)]X^{(B^+_uK^{*+},
\rho^0)}\right.\nonumber\\
&&\hspace{0.6cm}\left.+[V^*_{ub}V_{us}a^\lambda_1-V^*_{tb}V_{ts}
(a^\lambda_4+a^\lambda_{10})]X^{(B^+_u\rho^0,K^{*+})}\right\},
\\
\mathcal{A}^{SM}_\lambda(B^+_u\rightarrow \rho^+
K^{*0})&=&-\frac{G_F}{\sqrt{2}}V^*_{tb}V_{ts}
[a^\lambda_4-\frac{1}{2}a^\lambda_{10}]X^{(B^+_u\rho^+,K^{*0}
)},\\
\mathcal{A}^{SM}_\lambda(B^+_u\rightarrow \rho^0
\rho^+)&=&\frac{G_F}{2}\left\{V^*_{ub}V_{ud}\left[a^\lambda_1X^{(B^+_u\rho^0,
\rho^+)}+a^\lambda_2X^{(B^+_u\rho^+, \rho^0)}\right]\right.\nonumber\\
&&\hspace{0.6cm}\left.-V^*_{tb}V_{td}\left[\frac{3}{2}
\left(a^\lambda_7+a^\lambda_9+a^\lambda_{10}\right)X^{(B^+_u\rho^0,
\rho^+)}\right]\right\},\\
\mathcal{A}^{SM}_\lambda(B^0_d\rightarrow \rho^+
\rho^-)&=&\frac{G_F}{\sqrt{2}}
[V^*_{ub}V_{ud}a^\lambda_1-V^*_{tb}V_{td}
(a^\lambda_4+a^\lambda_{10})]X^{(B^0_d\rho^-,\rho^+)}.
\end{eqnarray}
where we neglected the annihilation matrix element contributions.

\section{\hspace{-0.6cm}. The Hamiltonian for RPV}
\vspace{-1cm}
\begin{eqnarray}
\mathcal{H}^{\spur{R}}_{eff}(B^+_u\rightarrow \phi K^{*+}
)&=&\frac{\lambda'_{i23}\lambda'^*_{i22}}
{2m^2_{\hat{\nu}_{Li}}}\eta^{-8/\beta_0}(\bar{s}\gamma^\mu
P_Ls)_8 (\bar{b}\gamma_\mu
P_Rs)_8+\frac{\lambda'_{i22}\lambda'^*_{i32}}
{2m^2_{\hat{\nu}_{Li}}}\eta^{-8/\beta_0}(\bar{s}\gamma^\mu
P_Rs)_8 (\bar{b}\gamma_\mu P_Ls)_8,\nonumber\\
\mathcal{H}^{\spur{R}}_{eff}(B^0_d\rightarrow \phi K^{*0}
)&=&\frac{\lambda'_{i23}\lambda'^*_{i22}}
{2m^2_{\hat{\nu}_{Li}}}\eta^{-8/\beta_0}(\bar{s}\gamma^\mu
P_Ls)_8 (\bar{b}\gamma_\mu
P_Rs)_8+\frac{\lambda'_{i22}\lambda'^*_{i32}}
{2m^2_{\hat{\nu}_{Li}}}\eta^{-8/\beta_0}(\bar{s}\gamma^\mu
P_Rs)_8 (\bar{b}\gamma_\mu P_Ls)_8, \nonumber\\
\mathcal{H}^{\spur{R}}_{eff}(B^+_u\rightarrow \rho^0 K^{*+}
)&=&\frac{\lambda''_{131}\lambda''^*_{121}}
{2\sqrt{2}m^2_{\hat{d}}}\eta^{-4/\beta_0}\left\{[(\bar{u}\gamma^\mu
P_Ru)_1 (\bar{b}\gamma_\mu P_Rs)_1-(\bar{u}\gamma^\mu P_Ru)_8
(\bar{b}\gamma_\mu P_Rs)_8]\right.\nonumber\\
&&\left. \hspace{3cm}-[(\bar{u}\gamma^\mu P_Rs)_1
(\bar{b}\gamma_\mu P_Ru)_1-(\bar{u}\gamma^\mu P_Rs)_8
(\bar{b}\gamma_\mu
P_Ru)_8]\right\}\nonumber\\
&&-\frac{\lambda''_{i31}\lambda''^*_{i21}}
{4\sqrt{2}m^2_{\hat{u}_{i}}}\eta^{-4/\beta_0}[(\bar{d}\gamma^\mu
P_Rd)_1 (\bar{b}\gamma_\mu P_Rs)_1-(\bar{d}\gamma^\mu P_Rd)_8
(\bar{b}\gamma_\mu
P_Rs)_8]\nonumber\\
&&+\left(\frac{\lambda'_{i13}\lambda'^*_{i12}}
{2\sqrt{2}m^2_{\hat{e}_{Li}}}(\bar{u}\gamma^\mu
P_Lu)_8 (\bar{b}\gamma_\mu
P_Rs)_8-\frac{\lambda'_{i13}\lambda'^*_{i12}}
{2\sqrt{2}m^2_{\hat{\nu}_{Li}}}(\bar{d}\gamma^\mu
P_Ld)_8 (\bar{b}\gamma_\mu
P_Rs)_8\nonumber\right.\\&&\left.-\frac{\lambda'_{i21}
\lambda'^*_{i31}}{2\sqrt{2}m^2_{\hat{\nu}_{Li}}}(\bar{d}\gamma^\mu
P_Rd)_8 (\bar{b}\gamma_\mu P_Ls)_8\right)\eta^{-8/\beta_0}, \nonumber\\
\mathcal{H}^{\spur{R}}_{eff}(B^+_u\rightarrow \rho^+ K^{*0} )&=&
\frac{\lambda''_{i31}\lambda''^*_{i12}}{4m^2_{\hat{u}_{i}}}
\eta^{-4/\beta_0}[(\bar{d}\gamma^\mu
P_Rs)_1 (\bar{b}\gamma_\mu P_Rd)_1-(\bar{d}\gamma^\mu P_Rs)_8
(\bar{b}\gamma_\mu
P_Rd)_8]\nonumber\\
&&\hspace{-1.5cm}+\frac{\lambda'_{i23}\lambda'^*_{i11}}{2m^2_{\hat{\nu}_{Li}}}
\eta^{-8/\beta_0}(\bar{d}\gamma^\mu
P_Ls)_8 (\bar{b}\gamma_\mu
P_Rd)_8+\frac{\lambda'_{i11}\lambda'^*_{i32}}
{2m^2_{\hat{\nu}_{Li}}}\eta^{-8/\beta_0}(\bar{d}\gamma^\mu
P_Rs)_8 (\bar{b}\gamma_\mu
P_Ld)_8,\nonumber\\
\mathcal{H}^{\spur{R}}_{eff}(B^+_u\rightarrow \rho^0 \rho^+
)&=&\frac{\lambda''_{132}\lambda''^*_{112}}
{2\sqrt{2}m^2_{\hat{s}}}\eta^{-4/\beta_0}\left\{[(\bar{u}\gamma^\mu
P_Ru)_1 (\bar{b}\gamma_\mu P_Rd)_1-(\bar{u}\gamma^\mu P_Ru)_8
(\bar{b}\gamma_\mu
P_Rd)_8]\right.\nonumber\\
&&\left.\hspace{2.8cm}- [(\bar{u}\gamma^\mu P_Rd)_1
(\bar{b}\gamma_\mu P_Ru)_1-(\bar{u}\gamma^\mu P_Rd)_8
(\bar{b}\gamma_\mu
P_Ru)_8]\right\}\nonumber\\
&&+\left(\frac{\lambda'_{i13}\lambda'^*_{i11}}
{2\sqrt{2}m^2_{\hat{e}_{Li}}}(\bar{u}\gamma^\mu
P_Lu)_8 (\bar{b}\gamma_\mu
P_Rd)_8-\frac{\lambda'_{i13}\lambda'^*_{i11}}
{2\sqrt{2}m^2_{\hat{\nu}_{Li}}}(\bar{d}\gamma^\mu
P_Ld)_8 (\bar{b}\gamma_\mu
P_Rd)_8\nonumber\right.\\&&\left.-\frac{\lambda'_{i11}
\lambda'^*_{i31}}{2\sqrt{2}m^2_{\hat{\nu}_{Li}}}
(\bar{d}\gamma^\mu
P_Rd)_8 (\bar{b}\gamma_\mu P_Ld)_8\right)\eta^{-8/\beta_0}, \nonumber\\
\mathcal{H}^{\spur{R}}_{eff}(B^0_d\rightarrow \rho^+ \rho^-
)&=&-\frac{\lambda''_{132}\lambda''^*_{112}}
{2m^2_{\hat{s}}}\eta^{-4/\beta_0}[(\bar{u}\gamma^\mu
P_Rd)_1 (\bar{b}\gamma_\mu P_Ru)_1-(\bar{u}\gamma^\mu P_Rd)_8
(\bar{b}\gamma_\mu P_Ru)_8],\nonumber
\end{eqnarray}

\section{\hspace{-0.6cm}. The amplitudes for RPV}
\vspace{-1cm}
\begin{eqnarray}
\mathcal{A}^{\spur{R}}_{\lambda}(B^+_u\rightarrow \phi K^{*+}
)&=&\frac{\lambda'_{i23}\lambda'^*_{i22}}{8N_Cm^2_{\hat{\nu}_{Li
}}}\eta^{-8/\beta_0}\Bigg[{1+\frac{\alpha_s}{4\pi}
\frac{C_F}{N_C}}\left(f^{'\lambda}_I(1)
+f'^\lambda_{II}(1)\right)\Bigg]X'^{(B^+_uK^{*+},
\phi)}\nonumber\\
&+&\frac{\lambda'_{i22}\lambda'^*_{i32}}
{8N_Cm^2_{\hat{\nu}_{Li}}}\eta^{-8/\beta_0}\Bigg[{1-\frac{\alpha_s}{4\pi}
\frac{C_F}{N_C}}\left(f^\lambda_I(-1)+f^\lambda_{II}(-1)\right)\Bigg]X^{(B^+_uK^{*+},
\phi)}
,\nonumber\\
\mathcal{A}^{\spur{R}}_{\lambda}(B^0_d\rightarrow \phi K^{*0}
)&=&\frac{\lambda'_{i23}\lambda'^*_{i22}}{8N_Cm^2_{\hat{\nu}_{Li}}}
\eta^{-8/\beta_0}\Bigg[{1+\frac{\alpha_s}{4\pi}
\frac{C_F}{N_C}}\left(f^{'\lambda}_I(1)
+f'^\lambda_{II}(1)\right)\Bigg]X'^{(B^+_uK^{*0},
\phi)}
\nonumber\\
&+&\frac{\lambda'_{i22}\lambda'^*_{i32}}
{8N_Cm^2_{\hat{\nu}_{Li}}}\eta^{-8/\beta_0}
\Bigg[{1-\frac{\alpha_s}{4\pi}
\frac{C_F}{N_C}}\left(f^\lambda_I(-1)+f^\lambda_{II}(-1)\right)
\Bigg]X^{(B^+_uK^{*0},
\phi)}, \nonumber\\
 \mathcal{A}^{\spur{R}}_{\lambda}(B^+_u\rightarrow \rho^0
K^{*+})&=&\frac{\lambda''_{131}\lambda''^*_{121}}{8\sqrt{2}
m^2_{\hat{d}}}\eta^{-4/\beta_0}\Bigg[1-\frac{1}{N_C}
-\frac{\alpha_s}{4\pi}
\frac{C_F}{N_C}\left(f^{'\lambda}_I(-1)+f'^\lambda_{II}(-1)\right)\Bigg]
X'^{(B^+_uK^{*+},\rho^0)}\nonumber\\
&&-\frac{\lambda''_{131}\lambda''^*_{121}}{8\sqrt{2}m^2_{\hat{d}}}
\eta^{-4/\beta_0}\Bigg[1-\frac{1}{N_C}-\frac{\alpha_s}{4\pi}
\frac{C_F}{N_C}\left(f^{'\lambda}_I(-1)+f'^\lambda_{II}(-1)\right)
\Bigg]X'^{(B^+_u\rho^0,K^{*+})}\nonumber\\
&&-\frac{\lambda''_{i31}\lambda''^*_{i21}}{16\sqrt{2}m^2_{\hat{u}_i}}
\eta^{-4/\beta_0}\Bigg[1-\frac{1}{N_C}-\frac{\alpha_s}{4\pi}
\frac{C_F}{N_C}\left(f^{'\lambda}_I(-1)+f'^\lambda_{II}(-1)\right)\Bigg]
X'^{(B^+_uK^{*+},\rho^0)}\nonumber\\
&&+\frac{\eta^{-8/\beta_0}}{N_C\sqrt{2}}\left(\frac{\lambda'_{i13}
\lambda'^*_{i12}}{8m^2_{\hat{e}_{Li}}}-\frac{\lambda'_{i13}
\lambda'^*_{i12}}{8m^2_{\hat{\nu}_{Li}}}\right)
\Bigg[{1+\frac{\alpha_s}{4\pi}
\frac{C_F}{N_C}}\left(f^{'\lambda}_I(1)+f'^\lambda_{II}(1)\right)\Bigg]
X'^{(B^+_uK^{*+},\rho^0)}\nonumber\\
&&-\frac{\lambda'_{i21}\lambda'^*_{i31}}{8N_C\sqrt{2}
m^2_{\hat{\nu}_{Li}}}\eta^{-8/\beta_0}\Bigg[1-\frac{\alpha_s}{4\pi}
\frac{C_F}{N_C}\left(f^\lambda_I(-1)
+f^\lambda_{II}(-1)\right)\Bigg]X^{(B^+_uK^{*+},\rho^0)}
, \nonumber\\
\mathcal{A}^{\spur{R}}_{\lambda}(B^+_u\rightarrow \rho^+
K^{*0})&=&
\frac{\lambda''_{i31}\lambda''^*_{i12}}{16m^2_{\hat{u}_i}}
\eta^{-4/\beta_0}\Bigg[1-\frac{1}{N_C}-\frac{\alpha_s}{4\pi}
\frac{C_F}{N_C}\left(f^{'\lambda}_I(-1)+f'^\lambda_{II}
(-1)\right)\Bigg]X'^{(B^+_u\rho^+,K^{*0})}\nonumber\\
&&+\frac{\lambda'_{i23}
\lambda'^*_{i11}}{8N_Cm^2_{\hat{\nu}_{Li}}}
\eta^{-8/\beta_0}\Bigg[{1+\frac{\alpha_s}{4\pi}
\frac{C_F}{N_C}}\left(f^{'\lambda}_I(1)
+f'^\lambda_{II}(1)\right)\Bigg]X'^{(B^+_u\rho^+,K^{*0})}\nonumber\\
&&+\frac{\lambda'_{i11}\lambda'^*_{i32}}{8N_Cm^2_{\hat{\nu}_{Li}}}
\eta^{-8/\beta_0}\Bigg[1-\frac{\alpha_s}{4\pi}
\frac{C_F}{N_C}\left(f^\lambda_I(-1)+f^\lambda_{II}(-1)\right)
\Bigg]X^{(B^+_u\rho^+,K^{*0})}
, \nonumber\\
\mathcal{A}^{\spur{R}}_{\lambda}(B^+_u\rightarrow \rho^0 \rho^+
)&=&\frac{\lambda''_{132}\lambda''^*_{112}}{8\sqrt{2}m^2_{\hat{s}}}
\eta^{-4/\beta_0}\Bigg[1-\frac{1}{N_C}-\frac{\alpha_s}{4\pi}
\frac{C_F}{N_C}\left(f^{'\lambda}_I(-1)+f'^\lambda_{II}(-1)\right)
\Bigg]X'^{(B^+_u\rho^+,\rho^0)}\nonumber\\
&&-\frac{\lambda''_{132}\lambda''^*_{112}}{8\sqrt{2}m^2_{\hat{s}}}
\eta^{-4/\beta_0}\Bigg[1-\frac{1}{N_C}-\frac{\alpha_s}{4\pi}
\frac{C_F}{N_C}\left(f^{'\lambda}_I(-1)+f'^\lambda_{II}(-1)\right)
\Bigg]X'^{(B^+_u\rho^0,\rho^+)}\nonumber\\
&+&\frac{\eta^{-8/\beta_0}}{N_C\sqrt{2}}\left(\frac{\lambda'_{i13}
\lambda'^*_{i11}}{8m^2_{\hat{e}_{Li}}}
-\frac{\lambda'_{i13}\lambda'^*_{i11}}{8m^2_{\hat{\nu}_{Li}}}\right)
\Bigg[1+\frac{\alpha_s}{4\pi}
\frac{C_F}{N_C}\left(f^{'\lambda}_I(1)+f'^\lambda_{II}(1)\right)
\Bigg]X'^{(B^+_u\rho^+,\rho^0)}\nonumber\\
&&-\frac{\lambda'_{i11}\lambda'^*_{i31}}{8N_C\sqrt{2}m^2_{\hat{\nu}_{Li}}}
\eta^{-8/\beta_0}\Bigg[1-\frac{\alpha_s}{4\pi}
\frac{C_F}{N_C}\left(f^\lambda_I(-1)+f^\lambda_{II}(-1)\right)\Bigg]
X^{(B^+_u\rho^+,\rho^0)}, \nonumber\\
\mathcal{A}^{\spur{R}}_{\lambda}(B^0_d\rightarrow \rho^+ \rho^-
)&=&-\frac{\lambda''_{132}\lambda''^*_{112}}{8m^2_{\hat{s}}}
\eta^{-4/\beta_0}\Bigg[1-\frac{1}{N_C}-\frac{\alpha_s}{4\pi}
\frac{C_F}{N_C}\left(f^{'\lambda}_I(-1)+f'^\lambda_{II}(-1)\right)
\Bigg]X'^{(B^0_d\rho^-,\rho^+)},\nonumber
\end{eqnarray}

\end{appendix}

\end{document}